\documentclass[reprint,tightenlines,eqsecnum,floats,nofootinbib,amsmath,amssymb,aps,prd]{revtex4-1}

\pdfoutput=1

\usepackage{amsfonts,amsthm,amscd}
\usepackage{graphicx}
\usepackage{enumerate} 
\usepackage{colordvi} 
\usepackage{units}
\usepackage{epsfig}
\usepackage{color}
\usepackage[english]{babel}
\usepackage[bookmarksopen]{hyperref}
\usepackage{tensor}
\usepackage{leftidx}
\usepackage[T1]{fontenc}
\usepackage{natbib}
\usepackage{undertilde}
\usepackage{mathtools}

\hypersetup{
pdfstartview = {FitH},
}

\newcommand{\ket}[1]{|#1\rangle}

\newcommand{\bk}[2]{\langle  #1|#2\rangle}

\newcommand{\comm}[2]{[#1,#2]}
\newcommand{\abs}[1]{|#1|}
\newcommand{\pb}[2]{\{ #1 , #2 \}}
\newcommand{\norm}[1]{\left\|#1\right\|}

\newcommand{\sslash}{\mathbin{/\mkern-6mu/}}

\DeclareMathOperator{\Tr}{Tr}
\DeclareMathOperator{\R}{Re}
\DeclareMathOperator{\I}{Im}
\DeclareMathOperator{\sl2c}{SL(2,\mathbb{C})}
\DeclareMathOperator{\asl2c}{\mathfrak{sl}(2,\mathbb{C})}
\DeclareMathOperator{\su2}{SU(2)}
\DeclareMathOperator{\asu2}{\mathfrak{su}(2)}
\DeclareMathOperator{\s11}{SU(1,1)}
\DeclareMathOperator{\as11}{\mathfrak{su}(1,1)}

\DeclareMathOperator{\diag}{diag}
\DeclareMathOperator{\arccot}{arccot}

\DeclareMathOperator{\bsl2c}{T^{\ast}SL(2,\mathbb{C})}
\DeclareMathOperator{\bsu2}{T^{\ast}SU(2)}
\DeclareMathOperator{\bs11}{T^{\ast}SU(1,1)}

\DeclarePairedDelimiter{\snorm}{\lVert}{\rVert}

\begin{document}
\title{\Large Timelike twisted geometries}
\author{Julian Rennert$^{1}$}
\email{jrennert@uwaterloo.ca}
\affiliation{$^1$ Department of Applied Mathematics, University of Waterloo, Waterloo, Ontario N2L 3G1, Canada}


\begin{abstract}
Within the twistorial parametrization of loop quantum gravity, we investigate the consequences of choosing a spacelike normal vector in the linear simplicity constraints. The amplitudes for the $\su2$ boundary states of loop quantum gravity, given by most of the current spin foam models, are constructed in such a way that even in the bulk only spacelike building blocks occur. Using a spacelike normal vector in the linear simplicity constraints allows us to distinguish spacelike from timelike 2-surfaces. We propose in this paper a quantum theory that includes both spatial and temporal building blocks and hence a more complete picture of quantum spacetime. At the classical level, we show how we can describe $\bs11$ as a symplectic quotient of 2-twistor space $\mathbb{T}^2$ by area matching and simplicity constraints. This provides us with the underlying classical phase space for $\s11$ spin networks describing timelike boundaries and their extension into the bulk. Applying a Dirac quantization, we show that the reduced Hilbert space is
spanned by $\s11$ spin networks and hence is able to give a quantum description of both spacelike and timelike faces. We discuss in particular the spectrum of the area operator and argue that for spacelike and timelike 2-surfaces it is discrete.
\end{abstract}


\maketitle


\section{Introduction}
\label{sec:intro}
Loop quantum gravity (LQG) is a canonical quantization of standard Einstein gravity in so-called connection variables and provides interesting insights into the nonperturbative structure of spatial quantum geometry \cite{rovelli, thiemann}. Spin foam models, on the other hand, aim at a covariant description of the same theory, using similar techniques. See, for example, Ref. \cite{rovellicov} for a recent introduction or Ref. \cite{perezsfreview}. The idea is that one can use a spin foam model to define a projector onto the physical Hilbert space of LQG by mapping kinematical spin network states onto states that solve the Hamiltonian constraint \cite{rovreis1,rovreis2,spinfoamprojector1}. The current Engle-Pereira-Rovelli-Livine-Freidel-Krasnov-Kaminski-Kisielowski-Lewandowski (EPRL-FK-KKL) spin foam model, named after the authors of Refs. \cite{spinfoam1,spinfoam2,spinfoam3,spinfoam5,spinfoam4}, solved several issues of its predecessors \cite{barrettcrane1,barrettcrane2,problemsbcsfm} such as having the correct boundary states to match the states of LQG and having a good semiclassical limit \cite{asymp1,asymp2,asymp3}. There are, however, further questions that are worth investigating. Possible improvements of the current spin foam model are discussed, for example, in Refs. \cite{linking1} and \cite{linking2} where the authors negate the question of whether the model defines a proper projector or rigging map onto the physical Hilbert space, and in Refs. \cite{proper1} and \cite{proper2}, the authors consider a modified vertex amplitude that improves the semiclassical limit compared to the original model. We would like to point out that it is possible that the work presented in this paper allows for an alternative approach to obtaining the results presented in Refs. \cite{proper1} and \cite{proper2}, not by restricting the vertex amplitude as in Refs. \cite{proper1} and \cite{proper2} but by generalizing it such that one sums over temporal building blocks as well as spatial ones. Further possible improvements of the current model are discussed also in Refs. \cite{cubulations} and \cite{secondary}.

The main motivation of this work, however, is related to the problem of timelike boundaries and the occurrence of nonspacelike building blocks in the bulk of spin foam models, which, in turn, relates to the study of timelike boundaries as motivated by the so-called general boundary formulation (GBF) \cite{oeckl1,oeckl2,oeckl3}. The absence of such nonspatial contributions in the current spin foam models was also discussed in Ref. \cite{immirzi}. Within the GBF, it is argued that, not only in quantum gravity but also in quantum theory in general, it is interesting, or even necessary, to consider amplitudes based on boundaries of finite regions of spacetime and to abandon the asymptotic states that are generally used in quantum field theory. These ideas are tightly connected to the framework of topological quantum field theory and constitute the basis for many considerations on amplitudes that are calculated from spin foam models. If we follow these ideas, we are led to the possibility of timelike boundaries and their corresponding amplitudes in Lorentzian spin foam models. In fact, the investigation of timelike components has a long history in this field \cite{timelikefaces1,timelikefaces2,timelikefaces3}.

Another motivation is to gain a better understanding of covariant quantum spacetime itself. If we consider spin foam models independently, \textit{a priori} not connected with LQG, can we use them to learn something about the quantum geometry of spacetime in the bulk? Currently, the new spin foam model is constructed in such a way that all its building blocks, even in the bulk, are strictly spacelike, which follows from the imposition of the linear simplicity constraints using a timelike normal vector $N^I$. This is necessary for achieving the matching of the spin foam boundary states with the kinematical $\su2$ spin network states of LQG. From a covariant standpoint, however, it is not clear why we should make such a restriction. Based on this reasoning, a generalization of the new spin foam model that uses both timelike as well as spacelike normal vectors $N^I$ for the linear simplicity constraints was proposed in Refs. \cite{LorentzianSFM1,LorentzianSFM2,unitaryirrep}. Their derivation is based on the Freidel-Krasnov model \cite{spinfoam5} and uses coherent states techniques to implement the simplicity constraints in the quantum theory. In this model one obtains an extra sum over the normal vector $N^I$ in the spin foam partition function, which can be understood in terms of the measure of the bivector field in the path integral as follows. In the standard case, where we only consider the timelike normal vector, the bivector fields are constrained to be spacelike and stay spacelike under gauge transformations. If we allow for a spacelike normal vector we can have spacelike, null and timelike bivectors and summing over timelike and spacelike $N^I$ can be justified by stating that we should integrate over all gauge-inequivalent contributions in the path integral. Now, this is certainly a statement about which dynamics is defined by the spin foam model. However, so far there has been no attempt at an asymptotic analysis of this generalized spin foam model.

This leads to our main objective for this work: namely, the question of whether we can give a twistorial description of the Conrady-Hnybida model \cite{LorentzianSFM1,LorentzianSFM2,unitaryirrep} with the hope that this would eventually allow for an asymptotic analysis of such generalized spin foam models with timelike components. In this paper we will first consider a phase space analysis in twistorial variables and leave the construction of a new spin foam model and its asymptotic analysis for future work.

The use of the twistorial parametrization of LQG \cite{twisted1,twisted2,twisted3,twisted4,twistphasespace} has in the past proven rather useful for the investigation of the covariance properties of LQG  \cite{covlqg,covlqg2} and the underlying phase space geometry. It has already been used in Ref. \cite{nulltwisted} to investigate the possibility of a null normal vector $N^I$ in the simplicity constraints and the subsequent quantization of null hypersurfaces with spacelike 2-surfaces. It also has recently been used to investigate conformal transformations in LQG \cite{miklos}. Very much in the same spirit of Ref. \cite{nulltwisted} we use these techniques here to consider timelike hypersurfaces with spacelike and timelike 2-surfaces. This can also be seen as a mathematical exercise further testing the adaptability of the twisted geometries formulation of LQG. We point out that, even though interesting by itself, our main interest here is not the (quantum) description of the spacelike but the timelike 2-surfaces. We find for example, similarly to the results obtained in Refs. \cite{geiller1} and \cite{geiller2} in a slightly different model, that the area spectrum of the timelike faces might be independent from the Barbero-Immirzi parameter.

One crucial question that has often been discussed in the literature on Lorentzian spin foam models is whether the (kinematical) spectra of geometrical operators are (all) discrete or continuous \cite{lengthandareaspec,lengthandareaspec2,lengthandareaspec3}. In 2+1 spacetime dimensions the situation is clear; see, for example, Ref. \cite{freidel3d} or the recent work \cite{sellaroli1,sellaroli2,sellaroli3}. There, one obtains continuous spectra for timelike 2-surfaces, because in that case the representations are labeled by a continuous parameter, which is a result of the noncompactness of the underlying gauge group. In 3+1 dimensions, however, the simplicity constraints can lead to relations between continuous and discrete representation labels, which amounts to the possibility that continuous spectra can become discrete. We will show that, indeed, also timelike faces can have discrete spectra when the simplicity constraints are imposed. This, however, requires a more detailed analysis than in the standard case with timelike $N^I$.

In the next section we review the description of $\bsl2c$ in terms of twistorial variables to fix our notation and conventions which are similar to those used in Ref. \cite{nulltwisted}. The difference with the original papers \cite{twisted2,twisted3,twisted4} is merely a sign flip in the Poisson brackets for the spinors. The Poisson structure we use here descends from the canonical one on twistor space \cite{tod}.

In section \ref{sec:timelike}, we investigate the symplectic reduction of 2-twistor space $\mathbb{T}^2$ by the simplicity and area matching constraints. As already mentioned, in the case of a spacelike normal vector $N^I$, we can have spacelike, timelike and null 2-surfaces. We focus here on the spacelike and timelike cases, which can be considered as being dual to each other. The phase space structure and symplectic reduction, in both cases, is very similar to the standard case with timelike normal, except that we obtain eventually $\bs11$ and not $\bsu2$.

In Sec. \ref{sec:fullgraph} we discuss general graphs. This requires us to impose the closure constraint in Sec. \ref{sec:closureconstraint} at the nodes, which is solved by $\s11$ intertwiners at the
nodes in the quantum theory.

Finally, we turn to the quantization in Sec. \ref{sec:quant}, which, again, proceeds similarly to the standard case. The difference
lies in the necessity to consider both half-links to obtain the full reduced Hilbert space, which is spanned by
all the unitary irreducible representations of $\s11$ that occur in the Plancherel decomposition. In Sec. \ref{sec:area}, we consider the area spectra associated with spacelike and timelike faces.

\section{Twistors in LQG and Spin foams}
\label{sec:twistors}
In this section, we want to give a brief overview of the utilization of spinors and twistors in loop gravity. Since their introduction in LQG and spin foams, see Refs. \cite{twisted1, twisted2, twisted3, twisted4, twistphasespace} and references therein, they have clarified many questions concerning the covariance properties of LQG as well as the relation between spin foams and the canonical theory. They provide a compelling picture for the spin network states of LQG as the quantization of certain (twisted) discrete geometries, and they have been used to investigate the quantization of null hypersurfaces in Ref. \cite{nulltwisted}.

In the current spin foam models, the starting point is the quantization of BF theory, on which one imposes the simplicity constraints, which reduce BF theory to general relativity, in the quantum theory. The BF action relates to the BF action with a Holst term and the Barbero-Immirzi parameter $\gamma \in \mathbb{R}_{\ast}$ through the so-called Immirzi shift and is given by
\begin{align}
S_{\text{BF}}[B,A] &= \int_M{\Tr\left(B \wedge F[A]\right)}\notag\\[0.3\baselineskip]
&= \int_M{\Tr\left(\ast \Sigma \wedge F[A] - \frac{1}{\gamma} \Sigma \wedge F[A]\right)}\,.
\label{eq:qu37}
\end{align}

\noindent
The $B$- and $\Sigma$- bivector fields take values in $\asl2c$, and $F[A]$ is the curvature of a $\asl2c$-valued spacetime connection $A$. The trace is taken with respect to the $\asl2c$ Cartan metric. The Immirzi shift amounts to a change of basis for $\asl2c$ in a way that leaves the equations of motion unaltered but changes the symplectic structure by introducing $\gamma$. BF theory is a topological theory and hence has only global degrees of freedom. By requiring that the $\Sigma$ field should be simple, i.e., $\Sigma^{IJ} = e^I \wedge e^J$, one obtains gravity (in the Einstein-Cartan form and up to a prefactor $1 / 16 \pi G$) with a Holst term \cite{holst}, i.e.,
\begin{equation}
S_{\text{Holst}}[e,A] = \int_M{\Tr\left(\ast e \wedge e \wedge F[A] - \frac{1}{\gamma} \, e \wedge e \wedge F[A]\right)}\,.
\label{eq:qu38}
\end{equation}

\noindent
In their linear form, those simplicity constraints are given by
\begin{equation}
N_{I} \Sigma^{IJ} = 0\,,
\end{equation}

\noindent
for some auxiliary normal vector $N^I$. Those constraints lead to two solutions, namely, $\Sigma^{IJ} = \pm e^I \wedge e^J$, where the sign relates to the orientation of the underlying frame field. This is relevant for the asymptotic analysis of the resulting spin foam model and has been investigated in Refs. \cite{proper1} and \cite{proper2}.

Using now a discretization of the spacetime manifold $M$ and a smearing of the continuous variables gives us a 2-complex decorated with $\bsl2c$ on each one-dimensional edge $e$ of the dual 2-complex. The group element $g$ corresponds to the holonomy of the connection $A$ along $e$ and can be used to measure the curvature associated with faces $f$ bounded by the edges $e_i$. The Lie algebra element corresponds to the smeared $B$ field over some 2-surface dual to $f$. We can now consider a three-dimensional intersection between this discrete structure and some hypersurface of spacetime. This leads us to some abstract, oriented graph $\Gamma$ with $N$ nodes $n$ and $L$ links $l$. Induced from the 2-complex, $\bsl2c$ is again associated with the links $l$. One reason for the name twisted geometries is the fact that $\bsl2c$ can be embedded in 2-twistor space as a symplectic quotient with respect to the so-called area matching constraint. Hence, we consider on each link a set of two twistors $(Z,W) \in \mathbb{T}^2 \cong \mathbb{C}^8$, where the first twistor is associated with the source node of the link and the second one is associated with the target node. Each twistor by itself is composed of two spinors $Z^{\alpha} = (\omega^A, i \bar{\pi}_{\bar{B}})$ and $W^{\alpha} = (\lambda^A, i \bar{\sigma}_{\bar{B}})$, where $\omega, \lambda  \in \mathbb{C}^2$ transforms under the $(\frac{1}{2},0)$ (left-handed) and $\bar{\pi}, \bar{\sigma} \in (\bar{\mathbb{C}}^2)^{\ast}$ transforms under the $(0,\frac{1}{2})$ (right-handed) representation of $\sl2c$. The adjoint twistors are given via $\bar{Z}_{\alpha} = (-i \pi_A,\bar{\omega}^{\bar{B}})$ such that the twistor norm is given by $\frac{1}{2} \bar{Z}_{\alpha} Z^{\alpha} = \I(\pi \omega)$. We use the convention $\epsilon^{01} = \epsilon_{01} = 1$, $\epsilon^{AB} = - \epsilon^{BA}$ for the two-dimensional $\epsilon$ tensor, which allows us to move spinor indices as
\begin{equation}
\omega^A = \epsilon^{AB} \omega_B \quad , \quad \omega_A = \epsilon_{BA} \omega^B
\label{eq:intro100}
\end{equation}

\noindent
and analogously for the complex conjugate sector. The 2-twistor space $\mathbb{T}^2$ comes equipped with a natural Poisson structure which is $\sl2c$ invariant and is given by the 2-form \cite{tod,penroserindler1,penroserindler2}\footnote{Following the conventions of the original twisted geometries literature \cite{twisted1,twisted2,twisted3,twisted4,nulltwisted}, we remove the $i$ appearing in the original spinorial Poisson brackets by parametrizing the twistors $Z$ and $W$ with an extra $i$ in front of $\bar{\pi}_{\bar{B}}$ and $\bar{\sigma}_{\bar{B}}$. As in Ref. \cite{nulltwisted}, we furthermore use the Poisson structure as defined by Eq.(\ref{eq:intro101}) and not with a relative minus sign. This leads to the symmetric Poisson brackets as shown in Eqs. (\ref{eq:intro103}) and (\ref{eq:intro103.5}).}
\begin{equation}
\Omega = i \, \text{d}Z^{\alpha} \wedge \text{d}\bar{Z}_{\alpha} + i \, \text{d}W^{\alpha} \wedge \text{d}\overline{W}_{\alpha}\,.
\label{eq:intro101}
\end{equation}

\noindent
In terms of the spinors Eq.(\ref{eq:intro101}) gives
\begin{align}
i \, \text{d}Z^{\alpha} \wedge \text{d}\bar{Z}_{\alpha} &= \text{d}\omega^{A} \wedge \text{d}\pi_A + \text{d} \bar{\omega}^{\bar{B}} \wedge \text{d} \bar{\pi}_{\bar{B}}\,,\label{eq:intro102}\\[0.3\baselineskip]
i \, \text{d}W^{\alpha} \wedge \text{d}\overline{W}_{\alpha} &= \text{d}\lambda^{A} \wedge \text{d}\sigma_A + \text{d} \bar{\lambda}^{\bar{B}} \wedge \text{d} \bar{\sigma}_{\bar{B}}\,,
\label{eq:intro102.5}
\end{align}

\noindent
which gives rise to the Poisson brackets
\begin{align}
\pb{\pi_A}{\omega^B} &= \delta^B_A = \pb{\sigma_A}{\lambda^B}\,,\label{eq:intro103}\\[0.3\baselineskip]
\pb{\bar{\pi}_{\bar{A}}}{\bar{\omega}^{\bar{B}}} &= \delta^{\bar{B}}_{\bar{A}} = \pb{\bar{\sigma}_{\bar{A}}}{\bar{\lambda}^{\bar{B}}}\label{eq:intro103.5}
\end{align}

\noindent
and all others vanishing. Thus, $\mathbb{T}^2$ together with the above brackets constitutes a Poisson manifold. For two functions $f,g$ on $\mathbb{T}^2$ we calculate their Poisson bracket via
\begin{align}
&\pb{f}{g} = \frac{\partial f}{\partial \pi_A} \frac{\partial g}{\partial \omega^A} - \frac{\partial f}{\partial \omega^A} \frac{\partial g}{\partial \pi_A} + \frac{\partial f}{\partial \sigma_A} \frac{\partial g}{\partial \lambda^A} - \frac{\partial f}{\partial \lambda^A} \frac{\partial g}{\partial \sigma_A}\notag\\[0.3\baselineskip]
&+ \frac{\partial f}{\partial \bar{\pi}_{\bar{A}}} \frac{\partial g}{\partial \bar{\omega}^{\bar{A}}} - \frac{\partial f}{\partial \bar{\omega}^{\bar{A}}} \frac{\partial g}{\partial \bar{\pi}_{\bar{A}}} + \frac{\partial f}{\partial \bar{\sigma}_{\bar{A}}} \frac{\partial g}{\partial \bar{\lambda}^{\bar{A}}} - \frac{\partial f}{\partial \bar{\lambda}^{\bar{A}}} \frac{\partial g}{\partial \bar{\sigma}_{\bar{A}}}\,.\label{eq:intro104}
\end{align}

\noindent
The \textit{area matching constraint}
\begin{equation}
C = \pi \omega - \lambda \sigma = 0
\label{eq:intro105}
\end{equation}

\noindent
is a first-class constraint and defines the embedding $\mathbb{T}^2_{\ast} \sslash C = \bsl2c$, \cite{twisted3,twisted4}. We assume throughout that $\pi \omega = \epsilon_{AB} \pi^A \omega^B = - \omega \pi \neq 0$ or $\sigma \lambda \neq 0$. Hence, we consider $\mathbb{T}^2_{\ast}$ where we remove the null configurations $\pi \omega = 0$ or $\sigma \lambda = 0$. One finds that the holonomy $g$ and the fluxes $\Pi$ of the gauge-invariant phase space $\bsl2c$ are parametrized in terms of the spinors via
\begin{equation}
\tensor[]{g}{^A_B} = \frac{\lambda^A \pi_B + \sigma^A \omega_B}{\sqrt{\pi \omega} \sqrt{\lambda \sigma}}\,,
\label{eq:intro106}
\end{equation}

\noindent
which satisfies $\det g = 1$ and $\pb{C}{\tensor[]{g}{^A_B}} = 0$, and
\begin{align}
\Pi^{AB} &= \frac{1}{4}\left(\pi^A \omega^B + \omega^A \pi^B\right) = \frac{1}{2} \pi^{(A} \omega^{B)}\,,\label{eq:intro110}\\[0.3\baselineskip]
\tilde{\Pi}^{AB} &= \frac{1}{4}\left(\sigma^A \lambda^B + \lambda^A \sigma^B\right) = \frac{1}{2} \sigma^{(A} \lambda^{B)}\,.\label{eq:intro110.5}
\end{align}

\noindent
Furthermore, one can show that
\begin{align}
&\pb{\tensor[]{g}{^A_B}}{\tensor[]{g}{^C_D}} = \label{eq:intro107}\\[0.3\baselineskip]
&\quad \frac{2 C}{(\pi \omega)^2 (\lambda \sigma)^2} \left[(\lambda \sigma) \epsilon^{AC} \Pi_{BD} - (\pi \omega) \epsilon_{BD} \tilde{\Pi}^{AC}\right]\notag
\end{align}

\noindent
and hence on the constraint surface $C = 0$, we get
\begin{equation}
\pb{\tensor[]{g}{^A_B}}{\tensor[]{g}{^C_D}} \approx 0\,.
\label{eq:intro108}
\end{equation}

\noindent
The group element $g$ defines a linear map from $Z$ to $W$:
\begin{align}
\tensor[]{g}{^A_B} \omega^B &= \sqrt{\frac{\pi \omega}{\lambda \sigma}} \, \lambda^A \approx \lambda^A\,,\label{eq:intro109}\\[0.3\baselineskip]
\tensor[]{g}{^A_B} \pi^B &= -\sqrt{\frac{\pi \omega}{\lambda \sigma}} \, \sigma^A \approx - \sigma^A\,.
\label{eq:intro109.5}
\end{align}

\noindent
A real bivector $B^{IJ}$ can be decomposed into a self-dual and an anti-self-dual part which, in spinorial variables, takes the following form:
\begin{equation}
B^{A\bar{B}C\bar{D}} = \Pi^{AC} \bar{\epsilon}^{\bar{B}\bar{D}} + \bar{\Pi}^{\bar{B}\bar{D}} \epsilon^{AC}\,.
\end{equation}

\noindent
Using
\begin{align}
\pb{C}{\omega^A} &= \omega^A \quad , \quad \pb{C}{\pi^A} = - \pi^A\,,\label{eq:intro111}\\[0.3\baselineskip]
\pb{C}{\lambda^A} &= \lambda^A \quad , \quad \pb{C}{\sigma^A} = - \sigma^A\label{eq:intro111.5}
\end{align}

\noindent
we show that $g, \Pi, \tilde{\Pi}$ are invariant under the flow of $C$. The fluxes transform like $\tilde{\Pi} \approx - g \Pi g^{-1}$ on the constraint surface $C=0$, and they furthermore satisfy two copies of the $\asl2c$ algebra,
\begin{align}
&\pb{\Pi^{AB}}{\Pi^{CD}} =\label{eq:intro113.1}\\[0.3\baselineskip]
&\quad \frac{1}{4} \left(\Pi^{AC} \epsilon^{BD} + \Pi^{AD} \epsilon^{BC} + \Pi^{BC} \epsilon^{AD} + \Pi^{BD} \epsilon^{AC}\right)\notag
\end{align}

\noindent
and similarly for the tilded fluxes, and we have
\begin{equation}
\pb{\Pi^{AB}}{\tilde{\Pi}^{CD}} = 0\,.
\label{eq:intro113.2}
\end{equation}

\noindent
Thus, the variables $g$ and $\Pi$ suffice to fully parametrize $\bsl2c$, and $\tilde{\Pi}$ is obtained from $g$ and $\Pi$ via $\tilde{\Pi} \approx - g \Pi g^{-1}$. We can now employ the following isomorphism between $\asl2c$ and $\mathbb{C}^3$ to rewrite the fluxes in terms of their rotation and boost generators according to
\begin{equation}
\tensor[]{\Pi}{^A_B} = \Pi^i \tensor[]{(\tau_i)}{^A_B} = \left(L^i + i \, K^i\right) \tensor[]{(\tau_i)}{^A_B}\,,
\label{eq:intro114}
\end{equation}

\noindent
with $i \in \{1,2,3\}$ and where the $\tau_i$ are related to the Pauli matrices via $\tau_i = \frac{1}{2i} \sigma_i$. They satisfy $[\tau_i,\tau_j] = \tensor[]{\varepsilon}{_i_j^k} \, \tau_k$, and we use them to calculate the components $\Pi^i \in \mathbb{C}$ via


\begin{equation}
\Pi^i = -2 \Tr(\Pi \, \tau_i) = -2 \, \tensor[]{\Pi}{^A_B} \tensor[]{(\tau_i)}{^B_A}\,,
\label{eq:intro121}
\end{equation}

\noindent
which gives
\begin{align}
\Pi^1 &= i (\Pi^{00} - \Pi^{11})\,,\label{eq:intro122}\\[0.3\baselineskip]
\Pi^2 &= - (\Pi^{00} + \Pi^{11})\,,\label{eq:intro123}\\[0.3\baselineskip]
\Pi^3 &= -2 i \Pi^{01}\,.\label{eq:intro124}
\end{align}

\noindent
Together with Eq.(\ref{eq:intro113.1}), this leads to
\begin{equation}
\pb{\Pi^i}{\Pi^j} = \tensor[]{\varepsilon}{^i^j_k} \Pi^k\,.
\label{eq:intro125}
\end{equation}

\noindent
Hence, on $C=0$, we reproduce the Poisson structure of $\bsl2c$ given by
\begin{align}
&\pb{\Pi^i}{\tensor[]{g}{^A_B}} = \tensor[]{g}{^A_C} \tensor[]{(\tau_i)}{^C_B}\,,\label{eq:intro126}\\[0.3\baselineskip]
&\pb{\tilde{\Pi}^i}{\tensor[]{g}{^A_B}} = - \tensor[]{(\tau_i)}{^A_C} \tensor[]{g}{^C_B}\,,\label{eq:intro126.2}\\[0.3\baselineskip]
&\pb{\tensor[]{g}{^A_B}}{\tensor[]{g}{^C_D}} \approx 0\,.\label{eq:intro126.5}
\end{align}

\noindent
Furthermore, note the 2-to-1 symmetry of the parametrization of the holonomy flux variables in terms of spinors; i.e., $g, \Pi, \tilde{\Pi}$ are invariant under
\begin{equation}
(\omega,\pi,\lambda,\sigma) \mapsto (\pi,\omega,\sigma,\lambda)\,.
\label{eq:intro127}
\end{equation}

\section{Twistorial description of timelike hypersurfaces}
\label{sec:timelike}
We will use the twistorial parametrization reviewed above to investigate the reduction of $\mathbb{T}^2_{\ast}$ by the linear simplicity constraints and the area matching constraint. But first, let us consider the bivector field $B \in \bigwedge^2 \mathbb{R}^{1,3} \otimes \asl2c$. In $\sl2c$ BF theory the $B$ field is valued in $\asl2c$ and hence can be expanded in terms of a $\asl2c$ basis. This means that we can express $B^{IJ}$ with the $\asl2c$ generators $L^i$ and $K^i$ as
\begin{equation}
B = \{B^{IJ}\} = \begin{pmatrix} 0 & K^1 & K^2 & K^3 \\ - K^1 & 0 & L^3 & -L^2 \\ - K^2 & -L^3 & 0 & L^1 \\ - K^3 & L^2 & - L^1 & 0 \end{pmatrix}
\label{eq:bfield1}
\end{equation}

\noindent
or, equivalently,
\begin{align}
K^i &= -K_i = B^{0i}\,,\\[0.3\baselineskip]
L^i &= L_i = (\ast B)^{0i} = \frac{1}{2} \tensor[]{\varepsilon}{^0^i_j_k} B^{jk}\,,
\end{align}

\noindent
where we used the Hodge star operator $\ast$, which satisfies $\ast^2 = -1$ in four dimensions with Lorentzian signature $(-,+,+,+)$. This gives furthermore
\begin{equation}
\{ \ast B^{IJ}\} = \begin{pmatrix} 0 & L^1 & L^2 & L^3 \\ - L^1 & 0 & -K^3 & K^2 \\ - L^2 & K^3 & 0 & -K^1 \\ - L^3 & -K^2 & K^1 & 0 \end{pmatrix}\,.
\label{eq:bfield2}
\end{equation}

\noindent
The two $\asl2c$-invariant Casimirs $C_1 = \vec{L}^2 - \vec{K}^2$ and $C_2 = -2 \, \vec{L} \cdot \vec{K}$ are obtained from $B^2 = \frac{1}{2} B_{IJ} B^{IJ} = - \vec{K}^2 + \vec{L}^2$ and $C_2 = \frac{1}{2} (\ast B)_{IJ} B^{IJ} = -2 \, ( L^1 K^1 + L^2 K^2 + L^3 K^2) = 2 K_i L^i = -2 L_i K^i$. Note that for the Lorentzian signature we have $(\ast B)^2 = - B^2$. Not surprisingly, this already shows the possibility of nondefinite bivectors in the case of a spacelike normal vector in the linear simplicity constraints. For the standard time gauge, where $N^I = (1,0,0,0)^t$, we have $B^{0i} = 0$ and hence see that $B$ is projected onto a Euclidean subspace with $(+,+,+)$ signature where we are only left with $B^2 > 0$ (we exclude the degenerate case of null bivectors in our considerations). If we choose the spacelike vector $N^I = (0,0,0,1)^t$, we deal with a subspace of signature $(+,-,-)$ and hence have, even after using the simplicity constraints, the possibility of bivectors with positive or negative areas. Let us also point out that in four spacetime dimensions every bivector can be written as the sum of two simple bivectors \cite{clifford}.

\subsection{Phase space structure and timelike simplicity constraints}
\label{sec:phasespace}
Using the Immirzi shift and identifying $B^{IJ}$ with the $\asl2c$ generators as in Eqs. (\ref{eq:bfield1}) and (\ref{eq:bfield2}), the linear simplicity constraints for spacelike normal $N^I = (0,0,0,1)^t$, i.e., $\Sigma^{3i} = 0$, become
\begin{equation}
L^3 = -\frac{1}{\gamma} K^3 \quad , \quad K^1 = \frac{1}{\gamma} L^1 \quad , \quad K^2 = \frac{1}{\gamma} L^2\,.
\end{equation}

\noindent
Using these constraints, we can already see that the $\asl2c$ Casimirs $C_1$ and $C_2$ reduce to
\begin{equation}
C_1 \longrightarrow (1-\gamma^2) \, Q_{\as11} \: , \quad C_2 \longrightarrow 2 \gamma \, Q_{\as11}\,,
\end{equation} 

\noindent
where the $\as11$ Casimir is given by $Q_{\as11} = (L^3)^2 - (K^1)^2 - (K^2)^2$. Following the procedure laid out in Refs. \cite{twisted4,twistphasespace,nulltwisted} we aim now for a decomposition of the constraints $\Sigma^{3i} = 0$ in their spinorial parametrization into a Lorentz-invariant part and a second part, specified by the little group of $N^I$. This has the advantage that the nature of those constraints becomes more transparent, which simplifies the phase space analysis as well as the quantization. We begin by rewriting $B^{IJ}$ in spinorial variables. The simplicity constraints become
\begin{equation}
n_{A \bar{B}} \Sigma^{A \bar{B} C \bar{D}} = 0
\label{eq:tsc21}
\end{equation}

\noindent
with
\begin{align}
&n_{A \bar{B}} = \epsilon_{CA} \epsilon_{\bar{D} \bar{B}} n^{C \bar{D}} = \frac{i}{\sqrt{2}} (\sigma_I)_{A \bar{B}} N^I\,,\label{eq:tsc22}\\[0.3\baselineskip]
&\Sigma^{A \bar{B} C \bar{D}} = - \frac{1}{2} (\sigma_I)^{A \bar{B}} (\sigma_J)^{C \bar{D}} \Sigma^{IJ}\,.\label{eq:tsc22.5}
\end{align}

\noindent
We use the following basis for the isomorphism between 4-vectors and anti-Hermitian matrices [note the extra factor of $i$ in Eq. (\ref{eq:tsc22})]:
\begin{align}
(\sigma_0)^{A \bar{B}} &= (\sigma_0)_{A \bar{B}} = \begin{pmatrix} 1 & 0 \\ 0 & 1 \end{pmatrix}\,,\label{eq:intro1122}\\[0.3\baselineskip]
(\sigma_1)^{A \bar{B}} &= - (\sigma_1)_{A \bar{B}} = \begin{pmatrix} 0 & 1 \\ 1 & 0 \end{pmatrix}\,,\label{eq:intro1123}\\[0.3\baselineskip]
(\sigma_2)^{A \bar{B}} &= (\sigma_2)_{A \bar{B}} = \begin{pmatrix} 0 & -i \\ i & 0 \end{pmatrix}\,,\label{eq:intro1124}\\[0.3\baselineskip]
(\sigma_3)^{A \bar{B}} &= - (\sigma_3)_{A \bar{B}} = \begin{pmatrix} 1 & 0 \\ 0 & -1 \end{pmatrix}\,.\label{eq:intro1125}
\end{align}

\noindent
Next, we decompose $B$ in terms of its self-dual and anti-self-dual components $\Pi$ and $\bar{\Pi}$ as
\begin{equation}
B^{A \bar{B} C \bar{D}} = B^{AC}_1 \epsilon^{\bar{B} \bar{D}} + \bar{B}^{\bar{B} \bar{D}}_2 \epsilon^{AC}\,,
\end{equation}

\noindent
where
\begin{align}
B^{A C}_1 &= -\frac{1}{2} \tensor[]{B}{^A ^{\bar{B}} ^C _{\bar{B}}} = B^{CA}_1\,,\\[0.3\baselineskip]
\bar{B}^{\bar{B}\bar{D}}_2 &= -\frac{1}{2} \tensor[]{B}{^A ^{\bar{B}} _A ^{\bar{D}}} = \bar{B}^{\bar{D}\bar{B}}_2\,.
\end{align}

\noindent
Note that for real bivectors we have $\: B_2 = \bar{B}_1 \:$; otherwise the self-dual and anti-self-dual parts are not complex conjugates of each other. Including the Immirzi shift, we have
\begin{align}
B &= B_1 \bar{\epsilon} + \bar{B}_1 \epsilon = \Pi \bar{\epsilon} + \bar{\Pi} \epsilon\notag\\[0.3\baselineskip]
&= (i \Sigma_1 - \frac{1}{\gamma} \Sigma_1) \bar{\epsilon} + (-i  \bar{\Sigma}_1 - \frac{1}{\gamma} \bar{\Sigma}_1) \epsilon
\label{eq:bfield300}
\end{align}

\noindent
and hence
\begin{align}
B_1 &= \Pi = (i  - \frac{1}{\gamma}) \Sigma_1\,,\\[0.3\baselineskip]
\Sigma_1 &= - \frac{i \gamma}{\gamma + i} \Pi\,.
\end{align}

\noindent
The difference in decomposing $B$ or $\ast B$ into self-dual and anti-self-dual components is an extra $i$ factor for the self-dual part and a $-i$ factor for the anti-self-dual part. This will be relevant for the distinction of spacelike and timelike 2-surfaces. Hence, we get for the linear simplicity constraints from Eq.(\ref{eq:tsc21})
\begin{equation}
n_{A \bar{B}} \left(- \frac{i \gamma}{\gamma + i} \, \Pi^{AC} \epsilon^{\bar{B} \bar{D}} + \frac{i \gamma}{\gamma - i} \, \bar{\Pi}^{\bar{B} \bar{D}} \epsilon^{AC}\right) = 0
\label{eq:tsc32}
\end{equation}

\noindent
and the dual constraint $N_I (\ast \Sigma)^{IJ} = 0$ gives
\begin{equation}
n_{A \bar{B}} \left(\frac{\gamma}{\gamma + i} \, \Pi^{AC} \epsilon^{\bar{B} \bar{D}} + \frac{\gamma}{\gamma - i} \, \bar{\Pi}^{\bar{B} \bar{D}} \epsilon^{AC}\right) = 0\,.
\label{eq:tsc32.5}
\end{equation}

\noindent
This distinction is important for the following reason. In order to split Eq.(\ref{eq:tsc32}) according to the decomposition used in Refs. \cite{twisted4}, \cite{twistphasespace}, and \cite{nulltwisted} into a Lorentz-invariant part and the part invariant under the little group, we use two linearly independent null vectors (one real and one complex), which are furthermore orthogonal to each other (there is nothing that forces us to use the same procedure, except its success in the timelike and null cases, and thus we prefer to stay as close as possible). Now, even though we are using the spacelike normal vector $N^I=(0,0,0,1)^t$, which projects onto a pseudo-Riemannian subspace and hence allows for bivectors with nondefinite norm, decomposing the simplicity constraint with respect to those null vectors always leads to subspaces where the bivectors have a definite norm. However, since we have seen that under the Hodge dual the bivector norm changes its sign, we can use this to distinguish the simplicity constraints for spacelike from those for timelike 2-surfaces. This essentially corresponds to the necessity of choosing another auxiliary vector $U^I$ to distinguish those two cases in the Conrady-Hnybida construction \cite{LorentzianSFM1,LorentzianSFM2}.

To be more explicit, we know that for a timelike normal vector $N^I$ the solutions to the simplicity constraints lead to positive definite bivectors because they lie in a subspace with Euclidean signature. Hence, we can conclude from $N_I \Sigma^{IJ} = 0$ that $\Sigma = \pm e_1 \wedge e_2$ with $\Sigma^2 > 0$ and hence $(\ast \Sigma)^2 < 0$ and, vice versa, we can conclude from $N_I (\ast \Sigma)^{IJ} = 0$ that $\ast \Sigma = \pm \tilde{e}_1 \wedge \tilde{e}_2$ with $(\ast \Sigma)^2 > 0$ and hence $(\Sigma)^2 < 0$. Now, for a spacelike normal $N^I$, we still obtain from $N_I \Sigma^{IJ} = 0$ that $\Sigma = \pm e_1 \wedge e_2$ but now this does not imply $\Sigma^2 > 0$ any longer (because we are in a space with Lorentzian signature). The question arises as to how we should distinguish whether $\Sigma$ is spacelike or timelike. Note that \textit{a priori} it should be possible to obtain spacelike as well as timelike solutions from one constraint, i.e., either $N_I \Sigma^{IJ} = 0$ or $N_I (\ast \Sigma)^{IJ} = 0$. However, for now, we will investigate the reduction of $\bsl2c$ by both constraints Eqs. (\ref{eq:tsc32}) and (\ref{eq:tsc32.5}) and discuss the results further in Sec. \ref{sec:area}.

Following again Refs. \cite{twisted4}, \cite{twistphasespace}, and \cite{nulltwisted}, we decompose Eqs. (\ref{eq:tsc32}) and (\ref{eq:tsc32.5}) by projecting them onto the two null vectors $\frac{i}{\sqrt{2}} \omega_C \bar{\omega}_{\bar{D}}$ (real) and $\frac{i}{\sqrt{2}} n_{C \bar{E}} \bar{\omega}^{\bar{E}} \bar{\omega}_{\bar{D}}$ (complex). Contracting Eq.(\ref{eq:tsc32}) with $\frac{i}{\sqrt{2}} \omega_C \bar{\omega}_{\bar{D}}$ gives us
\begin{equation}
\frac{\pi \omega}{\gamma + i} - \frac{\bar{\pi} \bar{\omega}}{\gamma - i} = 0
\label{eq:tsc34}
\end{equation}

\noindent
or equivalently
\begin{equation}
F_1 \equiv \R(\pi \omega) - \gamma \I(\pi \omega) = 0\,,
\label{eq:tsc34.5}
\end{equation}

\noindent
where we exclude cases where $\norm{\omega}^2 = - (\sigma_3)_{A \bar{B}} \omega^A \bar{\omega}^{\bar{B}} = \abs{\omega^0}^2 - \abs{\omega^1}^2 = 0$. This is the Lorentz-invariant constraint that one obtains for the time gauge, and hence it makes sense to associate it with spacelike bivectors. The contraction of Eq.(\ref{eq:tsc32}) with $\frac{i}{\sqrt{2}} n_{C \bar{E}} \bar{\omega}^{\bar{E}} \bar{\omega}_{\bar{D}}$ and assuming that $\norm{\omega}^2 \neq 0$ gives similarly the following complex constraint, which, due to the presence of the normal, is only invariant under the little group, which is in this case $\s11$:
\begin{equation}
F_2 = G_2 \equiv n^{A \dot{B}} \pi_A \bar{\omega}_{\dot{B}} = 0\,.
\label{eq:tsc36}
\end{equation}

\noindent
Applying the same procedure to Eq.(\ref{eq:tsc32.5}) only changes the Lorentz invariant constraint, and Eq.(\ref{eq:tsc36}) is valid for both cases. Hence, we have for the dual case the constraints Eq.(\ref{eq:tsc36}) together with
\begin{equation}
\frac{\pi \omega}{\gamma + i} + \frac{\bar{\pi} \bar{\omega}}{\gamma - i} = 0
\label{eq:tsc37}
\end{equation}

\noindent
or
\begin{equation}
G_1 \equiv \R(\pi \omega) + \frac{1}{\gamma} \I(\pi \omega) = 0\,,
\label{eq:tsc37.5}
\end{equation}

\noindent
as an equivalent set of constraints replacing Eq.(\ref{eq:tsc32.5}). Since they are dual to the first set, we interpret them as the ones corresponding to the timelike case\footnote{We mention that one can use a different decomposition of $N_I \Sigma^{IJ} = 0$ to obtain a different set of constraints by allowing that $\norm{\omega}^2$ can be zero. This might lead to the timelike sector with fixed $N_I \Sigma^{IJ} = 0$ and without the need to investigate $N_I (\ast \Sigma)^{IJ} = 0$. This new set of constraints includes $G_1$ as well, but the complex constraint $G_2$ is replaced by two real second-class constraints $E = (\sigma_3)^{A \bar{B}} \pi_A \bar{\pi}_{\bar{B}}$ and $F = (\sigma_3)_{A \bar{B}} \omega^A \bar{\omega}^{\bar{B}}$. We will leave this for future investigation and thank Wolfgang Wieland for this observation.}. A more direct way to see that this is the correct way to associate the $(F_1,F_2)$ with spacelike bivectors and $(G_1,G_2)$ with timelike bivectors is to consider the area form
\begin{align}
\mathcal{A}^2 &= \frac{1}{2} \left(\Sigma_1 \bar{\epsilon} + \bar{\Sigma}_1 \epsilon\right) \left(\Sigma_1 \bar{\epsilon} + \bar{\Sigma}_1 \epsilon\right)\label{eq:tscnew100}\\[0.3\baselineskip]
&= \left(-\frac{i \gamma}{\gamma + i}\right)^2 \Pi_{AC} \Pi^{AC} + \left(\frac{i \gamma}{\gamma - i}\right)^2 \bar{\Pi}_{\bar{B} \bar{D}} \bar{\Pi}^{\bar{B} \bar{D}}\notag\\[0.3\baselineskip]
&= \frac{\gamma^2}{8} \left(\frac{(\pi \omega)^2}{(\gamma + i)^2} + \frac{(\bar{\pi} \bar{\omega})^2}{(\gamma - i)^2}\right) = \frac{\gamma^2}{4} \, \R\left(\frac{(\pi \omega)^2}{(\gamma + i)^2}\right)\,.\notag
\end{align}

\noindent
One finds that the solutions of the simplicity constraint $F_1 = 0$, which are given by $\pi \omega = (\gamma + i) \, \mathcal{J}$, with $\mathcal{J} \in \mathbb{R}_{\ast}$, lead to a positive area,
\begin{equation}
\left.\mathcal{A}^2\right|_{F_1 = 0} = \frac{\gamma^2}{4} \mathcal{J}^2 > 0\,,
\label{eq:twist6}
\end{equation}

\noindent
whereas the solutions of $G_1 = 0$, which are given by $\pi \omega = i (\gamma + i) \, \mathcal{K}$, with $\mathcal{K} \in \mathbb{R}_{\ast}$, lead to a negative area,
\begin{equation}
\left.\mathcal{A}^2\right|_{G_1 = 0} = - \frac{\gamma^2}{4} \mathcal{K}^2 < 0\,.
\label{eq:twist7}
\end{equation}

\noindent
Note, that in both cases the area (squared) depends quadratically on $\gamma$. Since we only used $F_1$ in Eq. (\ref{eq:twist6}) and $G_1$ in Eq. (\ref{eq:twist7}), it is clear that this statement is independent of the choice between the other constraints, namely, whether we use $G_2$, or the ones suggested in footnote 2, where $G_2$ is replaced by $E$ and $F$. Furthermore, this suggests that also in the quantum theory the area spectra of spacelike and timelike areas should depend on $\gamma$.

\subsubsection{Spacelike faces}
\label{sec:spacephasespace}
We consider in this subsection the classical analysis of the constraints $F_1$, $F_2$ together with the area matching constraint $C$ from Eq.(\ref{eq:intro105}) and investigate the symplectic reduction $\mathbb{T}_{\ast} \sslash F_1 \sslash F_2$. We will also use the following version of $F_1$:
\begin{equation}
\mathring{F}_1 \equiv (\gamma - i) (\pi \omega) - (\gamma + i) (\bar{\pi} \bar{\omega}) = 0\,.
\label{eq:class3}
\end{equation}

\noindent
We first look for the classical solutions to the constraints $F_1$ and $F_2$. From twistor theory and the solutions of the simplicity constraints in the standard time gauge case, we know that the spinors are linearly dependent, and hence we are working with simple twistors, which are determined by a single spinor. This motivates to make the ansatz 
\begin{equation}
\pi_A = - \xi \, (\sigma_3)_{A \bar{B}} \bar{\omega}^{\bar{B}} \quad , \quad \xi \in \mathbb{C}_{\ast}
\label{eq:class11}
\end{equation}  

\noindent
and one finds that this indeed solves $G_2 = F_2 = 0$ for all $\xi \in \mathbb{C}_{\ast}$. Plugging our ansatz into $F_1 = 0$, we find with $\xi = r_{\xi} \exp(i \varphi_{\xi})$
\begin{equation}
F_1 = \norm{\omega}^2 r_{\xi} \left[\cos(\varphi_{\xi}) - \gamma \sin(\varphi_{\xi})\right] \stackrel{!}{=} 0\,,
\label{eq:class12}
\end{equation}

\noindent
where we have defined $\norm{\omega}^2 = - (\sigma_3)_{A \bar{B}} \omega^A \bar{\omega}^{\bar{B}} = \abs{\omega^0}^2 - \abs{\omega^1}^2$. Hence, we get
\begin{equation}
\varphi_{\xi} = \varphi(\gamma) = \arccot(\gamma) = \arctan \left(\frac{1}{\gamma}\right)\,.
\label{eq:class13}
\end{equation}

\noindent
We see that we can solve $F_1 = 0 = F_2$ by choosing
\begin{equation}
\pi_A = - r_{\xi} \, e^{i \varphi(\gamma)} \, (\sigma_3)_{A \bar{B}} \bar{\omega}^{\bar{B}} \quad , \quad r_{\xi} \in \mathbb{R}_{\ast}
\label{eq:class14}
\end{equation}

\noindent
and that $(r_{\xi}, \omega^A)$ span our five-dimensional solution space within $\mathbb{T}$, which has eight real dimensions. We have the system of constraints
\begin{align}
&\pb{\mathring{F}_1}{F_2} = - 2 \gamma F_2 \approx 0\,,\label{eq:class15}\\[0.3\baselineskip]
&\pb{\mathring{F}_1}{\bar{F}_2} = 2 \gamma \bar{F}_2 \approx 0\,,\label{eq:class16}\\[0.3\baselineskip]
&\pb{F_2}{\bar{F}_2} = - i \I(\pi \omega)\,,\label{eq:class17}
\end{align}

\noindent
and together with the area matching constraint, we have
\begin{equation}
\pb{\mathring{F}_1}{C} = 0 = \pb{\mathring{F}_1}{\bar{C}}
\label{eq:class18}
\end{equation}

\noindent
and
\begin{align}
&\pb{F_2}{C} = - \pb{F_2}{\bar{C}} = F_2 \approx 0\,,\label{eq:class19}\\[0.3\baselineskip]
&\pb{\bar{F}_2}{C} = - \pb{\bar{F}_2}{\bar{C}} = -\bar{F}_2 \approx 0\,.\label{eq:class20}
\end{align}

\noindent
Hence, we see that $F_1$ and $C$ are of first class and $F_2$ is of second class. On the fundamental spinors, $\mathring{F}_1$ generates the following transformations:
\begin{align}
&\pb{\mathring{F}_1}{\omega^A} = (\gamma - i) \, \omega^A\,,\label{eq:class21}\\[0.3\baselineskip]
&\pb{\mathring{F}_1}{\pi^A} = -(\gamma - i) \, \pi^A\,,\label{eq:class22}\\[0.3\baselineskip]
&\pb{\mathring{F}_1}{\bar{\omega}^{\bar{A}}} = -(\gamma + i) \, \bar{\omega}^{\bar{A}}\,,\label{eq:class23}\\[0.3\baselineskip]
&\pb{\mathring{F}_1}{\bar{\pi}^{\bar{A}}} = (\gamma + i) \, \bar{\pi}^{\bar{A}}\,.\label{eq:class24}
\end{align}

\noindent
Since $F_1$ is a first-class constraint, it generates gauge transformations, and we are interested in the gauge-invariant four-dimensional solution space. Consider the following bracket, with $\norm{\omega}^2 = -(\sigma_3)_{A \bar{B}} \omega^A \bar{\omega}^{\bar{B}}$, for which we have
\begin{equation}
\pb{\mathring{F}_1}{\norm{\omega}^{\alpha}} = - i \alpha \norm{\omega}^{\alpha}\,.
\label{eq:class25}
\end{equation} 

\noindent
Can we find an expression of $r_{\xi}$ in terms of $\omega^A$, in order to parametrize the reduced phase space? Note that
\begin{equation}
\pb{\mathring{F}_1}{\pi \omega} = 0\,.
\label{eq:class26}
\end{equation}

\noindent
If we use the solution Eq.(\ref{eq:class14}) and assume that $r_{\xi}$ is a function of $\omega^A$, we find with
\begin{equation}
\pi \omega = r_{\xi}(\omega^A) \, e^{i \varphi(\gamma)} \norm{\omega}^2
\label{eq:class27}
\end{equation}

\noindent
and Eq.(\ref{eq:class26}) that $r_{\xi}(\omega^A)$ must satisfy 
\begin{equation}
\pb{\mathring{F}_1}{r_{\xi}(\omega^A)} \stackrel{!}{=} 2 i \, r_{\xi}(\omega^A)\,.
\label{eq:class28}
\end{equation}

\noindent
From this, we conclude that
\begin{equation}
r_{\xi}(\omega^A) = \frac{N}{\norm{\omega}^2}
\label{eq:class29}
\end{equation}

\noindent
for some arbitrary numerical prefactor $N \in \mathbb{R}_{\ast}$. Hence, the four-dimensional reduced phase space (the symplectic quotient $\mathbb{T}\sslash F_1\sslash F_2$) can be parametrized by a single spinor. However, we know from Eq.(\ref{eq:class21}) that $\omega^A$ itself is not a gauge-invariant variable and hence not a good coordinate on the reduced phase space. Before we get to this point, let us choose $N$ such that
\begin{equation}
\pi \omega = (\gamma + i) \, \mathcal{J}
\label{eq:class30}
\end{equation}

\noindent
for some $\mathcal{J} \in \mathbb{R}_{\ast}$. This is achieved for
\begin{equation}
N = (\gamma + i) \, \mathcal{J} \, e^{- i \varphi(\gamma)} = \sqrt{1+\gamma^2} \, \mathcal{J}\,,
\label{eq:class31}
\end{equation}

\noindent
where we used that
\begin{align}
e^{i \varphi(\gamma)} &= \cos(\arccot(\gamma)) + i \sin(\arccot(\gamma))\notag\\[0.3\baselineskip]
&= \sqrt{\frac{\gamma + i}{\gamma - i}}
\label{eq:class32}
\end{align}

\noindent
and hence we get
\begin{equation}
\pi_A = - (\gamma + i) \, \mathcal{J} \, \frac{(\sigma_3)_{A \bar{B}} \bar{\omega}^{\bar{B}}}{\norm{\omega}^2}\,.
\label{eq:class33}
\end{equation}

\noindent
On the non-gauge-invariant solution space of $F_1$ and $F_2$, the variable $\mathcal{J}$ is given by
\begin{equation}
\mathcal{J} = \frac{\norm{\omega}^2 r_{\xi}}{\sqrt{1+\gamma^2}}
\label{eq:class34}
\end{equation}

\noindent
and hence
\begin{equation}
\pb{\mathring{F}_1}{\mathcal{J}} = 0\,.
\label{eq:class35}
\end{equation}

\noindent
Now, let us find the spinor that parametrizes the reduced phase space. Making the ansatz
\begin{equation}
z^A(\omega^B) = \sqrt{M} \, \frac{\omega^A}{\norm{\omega}^{\tau}}\,,
\label{eq:class36}
\end{equation}

\noindent
for some number $M$, and requiring that $\pb{\mathring{F}_1}{z^A} = 0$, gives
\begin{align}
\pb{\mathring{F}_1}{z^A} &= z^A \left[\gamma - i + i \tau\right] \stackrel{!}{=} 0\notag\\[0.3\baselineskip]
&\Leftrightarrow \quad \tau = i \gamma + 1\,.
\label{eq:class37}
\end{align}

\noindent
Furthermore, we have
\begin{equation}
\norm{z}^2 = -(\sigma_3)_{A \bar{B}} z^A \bar{z}^{\bar{B}} = M
\label{eq:class38}
\end{equation}

\noindent
and we will choose $M = 2 \mathcal{J}$. Note that $\mathcal{J}$ can be positive or negative, and if we wish to emphasize this point, we write $\varepsilon \mathcal{J}$ where we consider $\mathcal{J} > 0$ and $\varepsilon \in \{\pm 1\}$.

\subsubsection{Timelike faces}
\label{sec:timephasespace}
We consider now the symplectic reduction of $\mathbb{T}_{\ast}$ by the dual simplicity constraints Eq.(\ref{eq:tsc32.5}). We will use again the following expression for $G_1$:
\begin{equation}
\mathring{G}_1 \equiv (\gamma - i) (\pi \omega) + (\gamma + i) (\bar{\pi} \bar{\omega}) = 0\,.
\label{eq:class9}
\end{equation}

\noindent
To obtain the classical solutions of $G_1$ and $G_2$, we use now the ansatz
\begin{equation}
\pi_A = - i \, \zeta \, (\sigma_3)_{A \bar{B}} \bar{\omega}^{\bar{B}} \quad , \quad \zeta \in \mathbb{C}_{\ast}\,,
\label{eq:class39}
\end{equation}

\noindent
where we use the extra $i$ factor compared with the spacelike case and find that this solves $G_2 = 0$ for all $\zeta \in \mathbb{C}_{\ast}$. To solve $G_1 = 0$, we find that $\zeta = r_{\zeta} \exp(i \varphi_{\zeta})$ has to satisfy
\begin{equation}
G_1 = \frac{1}{\gamma} \norm{\omega}^2 r_{\zeta} \left[\cos(\varphi_{\zeta}) - \gamma \sin(\varphi_{\zeta})\right] \stackrel{!}{=} 0\,,
\label{eq:class40}
\end{equation}

\noindent
from which we get
\begin{equation}
\varphi_{\zeta} = \varphi_{\xi} = \varphi(\gamma) = \arccot(\gamma) = \arctan \left(\frac{1}{\gamma}\right)\,.
\label{eq:class41}
\end{equation}

\noindent
The fact that we obtain the same dependence of the phase and the Barbero-Immirzi parameter in the standard case Eq.(\ref{eq:class13}) as well as the dual case Eq.(\ref{eq:class41}) is a result of our $i$ factor, which we used in Eq.(\ref{eq:class39}). Thus, we see that we can solve $G_1 = 0 = G_2$ by choosing
\begin{equation}
\pi_A = -i \, r_{\zeta} \, e^{i \varphi(\gamma)} \, (\sigma_3)_{A \bar{B}} \bar{\omega}^{\bar{B}} \quad , \quad r_{\zeta} \in \mathbb{R}_{\ast},
\label{eq:class42}
\end{equation}

\noindent
and again $(r_{\zeta}, \omega^A)$ can be seen to span our five-dimensional solution space. The same procedure as in the spacelike case leads us the the gauge-invariant spinor variables. We have the relations between the simplicity constraints,
\begin{align}
&\pb{\mathring{G}_1}{G_2} = 2 i \, G_2 \approx 0\,,\label{eq:class43}\\[0.3\baselineskip]
&\pb{\mathring{G}_1}{\bar{G}_2} = -2 i \, \bar{G}_2 \approx 0\,,\label{eq:class44}\\[0.3\baselineskip]
&\pb{G_2}{\bar{G}_2} = \pb{F_2}{\bar{F}_2} = - i \I(\pi \omega)\label{eq:class45}
\end{align}

\noindent
and together with the area matching constraint we have
\begin{equation}
\pb{\mathring{G}_1}{C} = 0 = \pb{\mathring{G}_1}{\bar{C}}\,.
\label{eq:class46}
\end{equation}

\noindent
Because $G_2 = F_2$, the brackets with $C$ and $\bar{C}$ are equivalently given by Eqs. (\ref{eq:class19}) and (\ref{eq:class20}). $\mathring{G}_1$ acts with an extra minus sign on the complex conjugated spinors
\begin{align}
&\pb{\mathring{G}_1}{\omega^A} = (\gamma - i) \, \omega^A\,,\label{eq:class47}\\[0.3\baselineskip]
&\pb{\mathring{G}_1}{\pi^A} = -(\gamma - i) \, \pi^A\,,\label{eq:class48}\\[0.3\baselineskip]
&\pb{\mathring{G}_1}{\bar{\omega}^{\bar{A}}} = (\gamma + i) \, \bar{\omega}^{\bar{A}}\,,\label{eq:class49}\\[0.3\baselineskip]
&\pb{\mathring{G}_1}{\bar{\pi}^{\bar{A}}} = - (\gamma + i) \, \bar{\pi}^{\bar{A}}\,.\label{eq:class50}
\end{align}

\noindent
Hence, we find that the constraint structure is the same as in the spacelike case with $G_1$ and $C$ being of first class and $F_2$ being a complex second-class constraint. We consider again
\begin{equation}
\pb{\mathring{G}_1}{\norm{\omega}^{\alpha}} = \alpha \gamma \norm{\omega}^{\alpha}
\label{eq:class51}
\end{equation} 

\noindent
and ask whether we can find an expression of $r_{\zeta}$ in terms of $\omega^A$. Now, we use again that
\begin{equation}
\pb{\mathring{G}_1}{\pi \omega} = 0\,.
\label{eq:class52}
\end{equation}

\noindent
Using the the solution Eq.(\ref{eq:class42}) and the assumption that we can express $r_{\zeta}$ as a function of $\omega^A$, we find with
\begin{equation}
\pi \omega = i \, r_{\zeta}(\omega^A) \, e^{i \varphi(\gamma)} \norm{\omega}^2
\label{eq:class53}
\end{equation}

\noindent
and Eq.(\ref{eq:class52}) that $r_{\zeta}(\omega^A)$ must satisfy 
\begin{equation}
\pb{\mathring{G}_1}{r_{\zeta}(\omega^A)} \stackrel{!}{=} - 2 \gamma \, r_{\zeta}(\omega^A)\,.
\label{eq:class54}
\end{equation}

\noindent
From this, we conclude again that
\begin{equation}
r_{\zeta}(\omega^A) = \frac{\kappa}{\norm{\omega}^2}
\label{eq:class55}
\end{equation}

\noindent
for some arbitrary numerical prefactor $\kappa \in \mathbb{R}_{\ast}$. Now, we want to choose $\kappa$ such that
\begin{equation}
\pi \omega = i (\gamma + i) \, \mathcal{K}
\label{eq:class56}
\end{equation}

\noindent
for some $\mathcal{K} \in \mathbb{R}_{\ast}$ which is achieved for
\begin{equation}
\kappa = (\gamma + i) \, \mathcal{K} \, e^{- i \varphi(\gamma)} = \sqrt{1+\gamma^2} \, \mathcal{K}
\label{eq:class57}
\end{equation}

\noindent
and hence we get
\begin{equation}
\pi_A = -i \, (\gamma + i) \, \mathcal{K} \, \frac{(\sigma_3)_{A \bar{B}} \bar{\omega}^{\bar{B}}}{\norm{\omega}^2}\,.
\label{eq:class58}
\end{equation}

\noindent
On the non-gauge-invariant solution space of $G_1$ and $G_2$, the variable $\mathcal{K}$ is given by
\begin{equation}
\mathcal{K} = \frac{\norm{\omega}^2 r_{\zeta}}{\sqrt{1+\gamma^2}} = - \frac{i \norm{\omega}^2 r_{\xi}}{\sqrt{1+\gamma^2}} = -i \mathcal{J}
\label{eq:class59}
\end{equation}

\noindent
and hence
\begin{equation}
\pb{\mathring{G}_1}{\mathcal{K}} = 0\,.
\label{eq:class60}
\end{equation}

\noindent
The spinor that parametrizes the reduced phase space is again found by making the ansatz
\begin{equation}
y^A(\omega^B) = \sqrt{M} \, \frac{\omega^A}{\norm{\omega}^{\tau}}\,,
\label{eq:class61}
\end{equation}

\noindent
for some complex number $M$, and further requiring that $\pb{\mathring{G}_1}{y^A} = 0$ holds, which gives
\begin{align}
\pb{\mathring{G}_1}{y^A} &= y^A \left[- \gamma \tau + \gamma - i\right] \stackrel{!}{=} 0\notag\\[0.3\baselineskip]
&\Leftrightarrow \quad \tau = 1 - \frac{i}{\gamma}\,.
\label{eq:class62}
\end{align}

\noindent
Hence,
\begin{equation}
y^A(\omega^B) = \sqrt{M} \, \frac{\omega^A}{\norm{\omega}^{1 - i / \gamma}}
\label{eq:class63}
\end{equation}

\noindent
with
\begin{equation}
\norm{y}^2 = -(\sigma_3)_{A \bar{B}} y^A \bar{y}^{\bar{B}} = M\,.
\label{eq:class64}
\end{equation}

\noindent
Note that we choose the normalization of $y^A$ such that $M = 2 \gamma \mathcal{K}$, which is motivated by the simple form the Dirac bracket attains on the reduced phase space\footnote{This is a possible choice we can make. However, as we will discuss in Sec. \ref{sec:area}, it is worth keeping track of the fate of the Barbero-Immirzi parameter $\gamma$. Also cf. footnote 4.}. Note furthermore that in the standard timelike case one restricts $\mathcal{J}$ to be strictly positive, because in that case $\norm{z}^2 = \abs{z^0}^2 + \abs{z^1}^2 \geq 0$ and $\mathcal{J} = 0$ is ruled out since we assumed throughout that $\pi \omega \neq 0$. This restriction was used to get rid of a $\mathbb{Z}_2$ symmetry of the reduction of $\mathbb{T}^2_{\ast}$ to $\bsl2c$, i.e. Eq.(\ref{eq:intro127}), and we have the same symmetry present. In our case, however, the norm of $z^A$ and $y^A$ is not positive definite. Hence, if we want to focus on this nondefiniteness we can write $\varepsilon \mathcal{J}$ and $\varepsilon \mathcal{K}$, where $\varepsilon \in \{\pm 1\}$.

Now, we want to calculate the Dirac bracket of the reduced spinor with its complex conjugate. We need the Dirac bracket on the reduced space to take care of the second-class constraints $F_2 = G_2$ and $\bar{F}_2 = \bar{G}_2$. We use
\begin{equation}
z^A = \sqrt{2 \mathcal{J}} \frac{\omega^A}{\norm{\omega}^{i \gamma + 1}} = \sqrt{\frac{2 \pi \omega}{(\gamma + i)}} \frac{\omega^A}{\norm{\omega}^{i \gamma + 1}}
\label{eq:dirac1}
\end{equation}

\noindent
and
\begin{equation}
\utilde{z}^A = \sqrt{2 \mathcal{\utilde{\mathcal{J}}}} \frac{\lambda^A}{\norm{\lambda}^{i \gamma + 1}} = \sqrt{\frac{2 \sigma \lambda}{(\gamma + i)}} \frac{\lambda^A}{\norm{\lambda}^{i \gamma + 1}}
\label{eq:dirac2}
\end{equation}

\noindent
as coordinates on the reduced space $\mathbb{T}^2_{\ast} \sslash F \cong \mathbb{C}^2 \times \mathbb{C}^2$, where $F = \{F_1,F_2,\utilde{F}_1,\utilde{F}_2\}$, and 
\begin{equation}
y^A = \sqrt{2 \gamma \mathcal{K}} \frac{\omega^A}{\norm{\omega}^{1 - i/\gamma}} = \sqrt{\frac{2 \gamma \pi \omega}{(i \gamma - 1)}} \frac{\omega^A}{\norm{\omega}^{1 - i/\gamma}}
\label{eq:dirac3}
\end{equation}

\noindent
and
\begin{equation}
\utilde{y}^A = \sqrt{2 \gamma \utilde{\mathcal{K}}} \frac{\lambda^A}{\norm{\lambda}^{1 - i/\gamma}} = \sqrt{\frac{2 \gamma \sigma \lambda}{(i \gamma - 1)}} \frac{\lambda^A}{\norm{\lambda}^{1 - i/\gamma}}
\label{eq:dirac4}
\end{equation}

\noindent
as coordinates on the reduced space $\mathbb{T}^2_{\ast} \sslash G \cong \mathbb{C}^2 \times \mathbb{C}^2$, where $G = \{G_1,G_2,\utilde{G}_1,\utilde{G}_2\}$. Let us already note that the system of constraints $F$ or $G$ together with the area matching constraint is reducible, which means that after imposing $F=0$ or $G=0$ part of $C$ is already satisfied. Hence, the final step of the reduction is only with a reduced area matching constraint. Now, we calculate the Dirac bracket on $\mathbb{C}^2 \times \mathbb{C}^2$ via
\begin{align}
\pb{z^A}{\bar{z}^{\bar{B}}}_{\text{D}} &= \pb{z^A}{\bar{z}^{\bar{B}}} - \pb{z^A}{F_2} \, M^{-1}_{12} \, \pb{\bar{F}_2}{\bar{z}^{\bar{B}}}\notag\\[0.3\baselineskip]
&\quad - \pb{z^A}{\bar{F}_2} \, M^{-1}_{21} \, \pb{F_2}{\bar{z}^{\bar{B}}}\,.
\label{eq:dirac5}
\end{align}

\noindent
Together with
\begin{align}
M &= \begin{pmatrix} \pb{F_2}{F_2} & \pb{F_2}{\bar{F}_2} \\ \pb{\bar{F}_2}{F_2} & \pb{\bar{F}_2}{\bar{F}_2} \end{pmatrix} = i \I(\pi \omega) \begin{pmatrix} 0 & -1 \\ 1 & 0 \end{pmatrix}\notag\\[0.3\baselineskip]
&\Rightarrow \quad  M^{-1} = \frac{i}{\I(\pi \omega)} \begin{pmatrix} 0 & -1 \\ 1 & 0 \end{pmatrix}\,,
\label{eq:dirac6}
\end{align}

\noindent
we find
\begin{equation}
\pb{z^A}{\bar{z}^{\bar{B}}} = \frac{i}{2 \mathcal{J}} \, z^A \bar{z}^{\bar{B}}\,,
\label{eq:dirac7}
\end{equation}

\begin{equation}
\pb{z^A}{F_2} \approx - \frac{n^{A\bar{B}} \bar{z}_{\bar{B}}}{\norm{\omega}^{2 i \gamma}}\,,
\label{eq:dirac8}
\end{equation}

\begin{equation}
\pb{\bar{F}_2}{\bar{z}^{\bar{B}}} \approx \frac{\bar{n}^{\bar{B} C} z_{C}}{\norm{\omega}^{-2 i \gamma}}\,,
\label{eq:dirac9}
\end{equation}

\noindent
where Eqs. (\ref{eq:dirac8}) and (\ref{eq:dirac9}) hold weakly on $F_2$ and $\bar{F}_2$, respectively. Furthermore, with
\begin{equation}
\pb{z^A}{\bar{F}_2} = \pb{F_2}{\bar{z}^{\bar{B}}} = 0
\label{eq:dirac10}
\end{equation}

\noindent
we finally obtain
\begin{equation}
\pb{z^A}{\bar{z}^{\bar{B}}}_{\text{D}} \approx i (\sigma_3)^{A \bar{B}} \approx \pb{\utilde{z}^A}{\bar{\utilde{z}}^{\bar{B}}}_{\text{D}}\,,
\label{eq:dirac11}
\end{equation}

\noindent
where we used
\begin{equation}
n^{A\bar{B}} = \frac{i}{\sqrt{2}} (\sigma_3)^{A \bar{B}}\,.
\label{eq:dirac12}
\end{equation}

\noindent
Similarly, we find for the dual case with
\begin{equation}
\pb{y^A}{\bar{y}^{\bar{B}}} = \frac{i}{2 \gamma \mathcal{K}} \, y^A \bar{y}^{\bar{B}}\,,
\label{eq:dirac13}
\end{equation}

\begin{equation}
\pb{y^A}{G_2} \approx - \frac{n^{A\bar{B}} \bar{y}_{\bar{B}}}{\norm{\omega}^{-\frac{2 i}{\gamma}}}\,,
\label{eq:dirac14}
\end{equation}

\begin{equation}
\pb{\bar{G}_2}{\bar{y}^{\bar{B}}} \approx \frac{\bar{n}^{\bar{B} C} y_{C}}{\norm{\omega}^{\frac{2 i}{\gamma}}}\,,
\label{eq:dirac15}
\end{equation}

\noindent
where again Eqs. (\ref{eq:dirac14}) and (\ref{eq:dirac15}) hold weakly on $G_2$ and $\bar{G}_2$, respectively. And with
\begin{equation}
\pb{y^A}{\bar{G}_2} = \pb{G_2}{\bar{y}^{\bar{B}}} = 0\,,
\label{eq:dirac16}
\end{equation}

\noindent
we get\footnote{If we would have not put the extra $\gamma$ in the normalization of the reduced spinor in Eqs. (\ref{eq:dirac3}) and (\ref{eq:dirac4}), these two Dirac brackets would be given by $\pb{y^A}{\bar{y}^{\bar{B}}}_{\text{D}} \approx \frac{i}{\gamma} (\sigma_3)^{A \bar{B}} \approx \pb{\utilde{y}^A}{\bar{\utilde{y}}^{\bar{B}}}_{\text{D}}$.}
\begin{equation}
\pb{y^A}{\bar{y}^{\bar{B}}}_{\text{D}} \approx i (\sigma_3)^{A \bar{B}} \approx \pb{\utilde{y}^A}{\bar{\utilde{y}}^{\bar{B}}}_{\text{D}}\,.
\label{eq:dirac17}
\end{equation}

\noindent
In the standard case, using the time gauge, one obtains for the Dirac brackets of the reduced spinors the harmonic oscillator brackets where $(\sigma_3)^{A \bar{B}}$ is replaced by $(\sigma_0)^{A \bar{B}} = \delta^{A \bar{B}}$. In our case, instead, we find that we have an additional relative minus sign between brackets for the spinor components, which reflects the Lorentzian structure underlying our reduction. Furthermore, let us point out that those reduced brackets can be obtained equivalently as the Kirillov-Kostant-Souriau brackets \cite{kirillov} on the coadjoint orbits of $\s11$ for a timelike representative. We will further discuss this point in Sec. \ref{sec:reducarea}. Before that, however, we will consider again the second-class constraints $F_2 = G_2$ and show that it can be exchanged for an equivalent real first-class constraint, the so-called master constraint, which will be important for the quantum theory, where it is easier to impose the first-class constraints strongly than properly taking care of the second class constraints.
We follow again the procedure known from the standard time-gauge case, where the first-class master constraint is defined via (equivalently for $G_2$)
\begin{equation}
\textbf{M} \equiv \bar{F}_2 F_2 = 0\,.
\label{eq:tsc2}
\end{equation}

\noindent
We can now rewrite $\textbf{M}$ in terms of quantities that simplify the identification of the solution space to $\textbf{M} = 0$ in the quantum theory. This is achieved by the fact that we can rewrite it in terms of one of the $\asl2c$ Casimirs and the $\as11$ Casimir plus an extra term, and for all of those, we know the spectrum on the noncanoncial basis of $\sl2c$, which diagonalizes not $\su2$ but $\s11$. We follow Ref. \cite{twistphasespace} closely and adapt it to the timelike case. We have
\begin{align}
&\textbf{M} = \bar{F}_2 F_2 = \bar{n}^{\dot{A} B} n^{C \dot{D}} \bar{\pi}_{\dot{A}} \omega_{B} \pi_{C} \bar{\omega}_{\dot{D}}\label{eq:tsc3}\\[0.5\baselineskip]
&= \bar{n}^{\dot{A} B} n^{C \dot{D}} \left(\omega_{(B} \pi_{C)} + \omega_{[B} \pi_{C]}\right) \left(\bar{\pi}_{(\dot{A}} \bar{\omega}_{\dot{D})} + \bar{\pi}_{[\dot{A}} \bar{\omega}_{\dot{D}]}\right)\,,\notag
\end{align}

\noindent
where we used that $\omega_{B} \pi_{C} = \left(\omega_{(B} \pi_{C)} + \omega_{[B} \pi_{C]}\right)$\,. We obtain
\begin{align}
\textbf{M} =& \, \bar{n}^{\dot{A} B} n^{C \dot{D}} \left(2 \, \Pi_{BC} + (\omega \pi) \, \epsilon_{BC}\right) \left(2 \, \bar{\Pi}_{\dot{A} \dot{D}} + (\bar{\pi} \bar{\omega}) \, \epsilon_{\dot{A} \dot{D}}\right)\notag\\[0.3\baselineskip]
=& \, \bar{n}^{\dot{A} B} n^{C \dot{D}} \left(4 \, \Pi_{BC} \bar{\Pi}_{\dot{A} \dot{D}} + 2 \, (\bar{\pi} \bar{\omega}) \, \Pi_{BC} \, \epsilon_{\dot{A} \dot{D}}\right.\notag\\[0.3\baselineskip]
&\left. \, + 2 \, (\omega \pi) \, \bar{\Pi}_{\dot{A} \dot{D}} \, \epsilon_{BC} - \abs{\pi \omega}^2 \epsilon_{BC} \epsilon_{\dot{A} \dot{D}}\right)\,.
\label{eq:tsc5}
\end{align}

\noindent
Together with $N^I=(0,0,0,1)$ and $n^{A \dot{B}} = \frac{i}{\sqrt{2}} (\sigma_I)^{A \dot{B}} N^I = \frac{i}{\sqrt{2}} \diag(1,-1)$, one can now show explicitly that
\begin{align}
\textbf{M} &= 4 \, \bar{n}^{\dot{A} B} n^{C \dot{D}} \, \Pi_{BC} \bar{\Pi}_{\dot{A} \dot{D}} - \abs{\pi \omega}^2 \bar{n}^{\dot{A} B} n^{C \dot{D}} \, \epsilon_{BC} \epsilon_{\dot{A} \dot{D}}\notag\\[0.3\baselineskip]
&= 4 \, \bar{n}^{\dot{A} B} n^{C \dot{D}} \, \Pi_{BC} \bar{\Pi}_{\dot{A} \dot{D}} + \abs{\pi \omega}^2
\label{eq:tsc9}
\end{align}

\noindent
For the first term in Eq.(\ref{eq:tsc9}), we get
\begin{equation}
4 \, \bar{n}^{\dot{A} B} n^{C \dot{D}} \, \Pi_{BC} \bar{\Pi}_{\dot{A} \dot{D}} = 2 \, \abs{\Pi_{00}}^2 - 4 \, \abs{\Pi_{01}}^2 + 2 \, \abs{\Pi_{11}}^2\,.
\label{eq:tsc10}
\end{equation}

\noindent
Let us now rewrite the fluxes in terms of their rotation and boost generators using Eqs. (\ref{eq:intro114}) and (\ref{eq:intro121}), which gives us
\begin{align}
&\abs{\Pi_{00}}^2 = \abs{\Pi_{11}}^2 = \frac{1}{4} \left(\abs{\Pi^1}^2 + \abs{\Pi^2}^2\right)\notag\\[0.3\baselineskip]
&= \frac{1}{4} \left((L^1)^2 + (L^2)^2 + (K^1)^2 + (K^2)^2\right)
\label{eq:tsc16}
\end{align}

\noindent
and
\begin{equation}
\abs{\Pi_{01}}^2 = \frac{1}{4} \abs{\Pi^3}^2 = \frac{1}{4} \left((L^3)^2 + (K^3)^2\right)\,.
\label{eq:tsc17}
\end{equation}

\noindent
Hence, we finally get for Eq.(\ref{eq:tsc10})
\begin{align}
&4 \, \bar{n}^{\dot{A} B} n^{C \dot{D}} \, \Pi_{BC} \bar{\Pi}_{\dot{A} \dot{D}}\label{eq:tsc19}\\[0.3\baselineskip]
&\quad = \left[(\vec{L})^2 - (\vec{K})^2 - 2 \left((L^3)^2 - (K^1)^2 - (K^2)^2\right)\right]\,.\notag
\end{align}

\noindent
Now, we note that $(\vec{L})^2 - (\vec{K})^2$ is the quadratic $\asl2c$ Casimir and furthermore $Q_{\as11} = (L^3)^2 - (K^1)^2 - (K^2)^2$ is the Casimir of $\as11$, and we get for the master constraint for a spacelike normal $N^I$
\begin{equation}
\textbf{M} = \left(C_{\sl2c} - 2 \, Q_{\as11}\right) + \abs{\pi \omega}^2\,.
\label{eq:tsc20}
\end{equation}

\noindent
Recall that for the case of timelike normal vector we obtain the $\asu2$ Casimir instead of $Q_{\as11}$, but otherwise it looks exactly the same. Finding the complete solution space in the quantum theory, however, is more involved than in the standard case.

\subsection{Reduction by the area matching constraint}
\label{sec:reducarea}
As we have mentioned before, the system of all constraints is reducible. On $\mathbb{T}^2_{\ast} \sslash F$ or $\mathbb{T}^2_{\ast} \sslash G$ part of the area matching constraint $C$ is already satisfied. One finds that the reduced area matching constraint is given by
\begin{equation}
C_{\text{red}} = \snorm[\big]{z}^2 + \snorm[\big]{\utilde{z}}^2 = 0
\label{eq:reduc1}
\end{equation}

\noindent
or in the dual case by 
\begin{equation}
D_{\text{red}} = \snorm[\big]{y}^2 + \snorm[\big]{\utilde{y}}^2 = 0\,.
\label{eq:reduc2}
\end{equation}

\noindent
Note that this constraint has nontrivial solutions, since the ``norm'' of the spinors $\norm{z}^2$, etc., is not positive definite in our case. We will see that these constraints will be solved by $\mathcal{J} = - \utilde{\mathcal{J}}$ and $\mathcal{K} = - \utilde{\mathcal{K}}$. We will use $\mathcal{J},\mathcal{K},\utilde{\mathcal{J}},\utilde{\mathcal{K}} > 0$ and solve the constraints by using opposite $\varepsilon$'s. Equivalently, we could have chosen the normalization of the tilded sector to be $M = -2 \utilde{\mathcal{J}}$ to obtain a reduced area matching with a minus sign, which was used in Refs. \cite{twisted1,twisted2,twisted3}. However, the important point is the gauge transformations that are generated by $C_{\text{red}}$ and $D_{\text{red}}$, and those are not affected by this sign. The origin of this minus sign can be traced back to our choice to have the standard Poisson structure on $\mathbb{T}^2$ and not the sign-flipped one used, for example, in Refs. \cite{twisted1,twisted2,twisted3,twisted4,twisted5}.

We are now interested in the reductions $(\mathbb{C}^2 \times \mathbb{C}^2) \sslash C_{\text{red}}$ and $(\mathbb{C}^2 \times \mathbb{C}^2) \sslash D_{\text{red}}$ and whether we end up with $\bs11$ in both cases. Remember that from now on we are using the Dirac bracket on the reduced phase space. We have
\begin{equation}
\pb{C_{\text{red}}}{z^A} = - i z^A \quad , \quad \pb{C_{\text{red}}}{\utilde{z}^A} = - i \utilde{z}^A\,,
\label{eq:reduc3}
\end{equation}
\begin{equation}
\pb{C_{\text{red}}}{\bar{z}^{\bar{A}}} = i \bar{z}^{\bar{A}} \quad , \quad \pb{C_{\text{red}}}{\bar{\utilde{z}}^{\bar{A}}} = i \bar{\utilde{z}}^{\bar{A}}
\label{eq:reduc3.1}
\end{equation}

\noindent
and similarly
\begin{equation}
\pb{D_{\text{red}}}{y^A} = - i y^A \quad , \quad \pb{D_{\text{red}}}{\utilde{y}^A} = - i \utilde{y}^A\,,
\label{eq:reduc4}
\end{equation}
\begin{equation}
\pb{D_{\text{red}}}{\bar{y}^{\bar{A}}} = i \bar{y}^{\bar{A}} \quad , \quad \pb{D_{\text{red}}}{\bar{\utilde{y}}^{\bar{A}}} = i \bar{\utilde{y}}^{\bar{A}}\,.
\label{eq:reduc4.1}
\end{equation}

\noindent
Inspired by the holonomy and the fluxes constructed in Sec. \ref{sec:twistors} we find that we can analogously parametrize the gauge-invariant reduced phase space $(\mathbb{C}^2 \times \mathbb{C}^2) \sslash C_{\text{red}}$ with the holonomy
\begin{equation}
\tensor[]{h}{^A_B} = \frac{\utilde{z}^A (\sigma_3)_{B \bar{C}} \bar{z}^{\bar{C}} + (\sigma_3)^{A\bar{C}} \utilde{\bar{z}}_{\bar{C}} z_B}{\snorm{z} \snorm{\utilde{z}}}
\label{eq:reduc5}
\end{equation}

\noindent
and similarly for $(\mathbb{C}^2 \times \mathbb{C}^2) \sslash D_{\text{red}}$,
\begin{equation}
\tensor[]{h}{^A_B} = \frac{\utilde{y}^A (\sigma_3)_{B \bar{C}} \bar{y}^{\bar{C}} + (\sigma_3)^{A\bar{C}} \utilde{\bar{y}}_{\bar{C}} y_B}{\snorm{y} \snorm{\utilde{y}}}\,.
\label{eq:reduc5.5}
\end{equation}

\noindent
They are both of the form
\begin{equation}
h = \begin{pmatrix} a & b \\ \bar{b} & \bar{a} \end{pmatrix}\,.
\end{equation}

\noindent
For Eq.(\ref{eq:reduc5}) we have
\begin{equation}
a = \frac{(z^1 \bar{\utilde{z}}^{\bar{1}} - \bar{z}^{\bar{0}} \utilde{z}^{0})}{\snorm{z} \snorm{\utilde{z}}} \quad , \quad b = \frac{(\utilde{z}^{0} \bar{z}^{\bar{1}}  - \bar{\utilde{z}}^{\bar{1}} z^0)}{\snorm{z} \snorm{\utilde{z}}}
\end{equation}

\noindent
and similarly for Eq.(\ref{eq:reduc5.5}) and thus both satisfy $\det h = 1$. Hence, we see that, indeed, we obtain $\s11$ on the reduced phase space. Furthermore, on $C_{\text{red}}$, we have
\begin{equation}
\utilde{z}^A \approx \tensor[]{h}{^A_B} z^B \quad , \quad \utilde{y}^A \approx \tensor[]{h}{^A_B} y^B
\label{eq:reduc6}
\end{equation}

\noindent
and one shows explicitly that, using the Dirac bracket, we have
\begin{equation}
\pb{C_{\text{red}}}{\tensor[]{h}{^A_B}} = 0
\label{eq:reduc6.1}
\end{equation}

\noindent
and
\begin{equation}
\pb{\tensor[]{h}{^A_B}}{\tensor[]{h}{^C_D}} \approx 0\,.
\label{eq:reduc6.2}
\end{equation}

\noindent
The fluxes $\Pi^{BD}$ from Eq.(\ref{eq:intro110}) become
\begin{equation}
\pi^{BD} = \frac{(\gamma + i)}{8} \left[(\sigma_3)^{B \bar{C}} \bar{z}_{\bar{C}} z^D + (\sigma_3)^{D \bar{C}} \bar{z}_{\bar{C}} z^B\right]
\label{eq:reduc7}
\end{equation}

\noindent
which gives
\begin{equation}
\pi = -\frac{(\gamma + i)}{8}\begin{pmatrix} 2 z^0 \bar{z}^{\bar{1}} & (\abs{z^0}^2 + \abs{z^1}^2) \\ (\abs{z^0}^2 + \abs{z^1}^2) & 2 \bar{z}^{\bar{0}} z^1 \end{pmatrix}\,.
\label{eq:reduc8}
\end{equation}

\noindent
They satisfy, of course,
\begin{equation}
\pb{C_{\text{red}}}{\pi^{BD}} = 0\,.
\end{equation}

\noindent
We can now expand $\pi$ in terms of a $\as11$ basis, i.e., $\pi = \pi^i \tau_i$. With
\begin{align}
\tensor[]{(\tau_1)}{^A_B} &= \frac{1}{2 i} \begin{pmatrix} 1 & 0 \\ 0 & -1 \end{pmatrix}\,,\notag\\[0.3\baselineskip]
\tensor[]{(\tau_2)}{^A_B} &= \frac{1}{2} \begin{pmatrix} 0 & 1 \\ 1 & 0 \end{pmatrix}\,,\label{eq:reduc8.5}\\[0.3\baselineskip]
\tensor[]{(\tau_3)}{^A_B} &= \frac{1}{2} \begin{pmatrix} 0 & -i \\ i & 0 \end{pmatrix}\notag
\end{align}

\noindent
and by a rescaling with a factor $-2i / (\gamma + i)$, we get
\begin{align}
\pi^1 &= \frac{1}{2}(\abs{z^0}^2 + \abs{z^1}^2)\,,\label{eq:reduc9}\\[0.3\baselineskip]
\pi^2 &= \I(\bar{z}^{\bar{0}} z^1)\,,\label{eq:reduc10}\\[0.3\baselineskip]
\pi^3 &= -\R(\bar{z}^{\bar{0}} z^1)\,.\label{eq:reduc11}
\end{align}

\noindent
They satisfy
\begin{equation}
\pb{\pi^1}{\pi^3} = \pi^2\,,
\end{equation}
\begin{equation}
\pb{\pi^1}{\pi^2} = -\pi^3\,,
\end{equation}
\begin{equation}
\pb{\pi^3}{\pi^2} = -\pi^1\,,
\end{equation}

\noindent
and hence we see that we get indeed a $\as11$ algebra where $(\pi^1,\pi^2,\pi^3) \cong (J_3,K_2,K_1)$. Thus, we see that we finally obtain $\bs11$ via a symplectic reduction of $\mathbb{T}^2_{\ast}$ by the simplicity constraints and the area-matching constraint. This holds in both cases of constraints $(F,C)$ and $(G,C)$. In terms of the reduced spinors, one finds that with Eqs. (\ref{eq:reduc9}) - (\ref{eq:reduc11}) the $\as11$ Casimir operator is given by
\begin{align}
Q_{\as11} &= (\pi^1)^2 - (\pi^2)^2 - (\pi^3)^2\\[0.3\baselineskip]
&= \frac{1}{4} (\abs{z^0}^2 - \abs{z^1}^2) = \frac{1}{4} \snorm{z}^2\,.
\end{align}

\noindent
Now, as we have mentioned before, let us show that the Poisson structure we have obtained via reduction from $\mathbb{T}^2_{\ast}$ by the simplicity and area matching constraint is exactly the canonical symplectic structure (Kirillov-Kostant-Souriau symplectic structure \cite{kirillov}) on the coadjoint orbits of $\s11$. If we take an element $g \in \s11$ with
\begin{equation}
g = \begin{pmatrix} z^0 & z^1 \\ \bar{z}^{\bar{1}} & \bar{z}^{\bar{0}} \end{pmatrix} \quad , \quad \abs{z^0}^2 - \abs{z^1}^2 = 1\,.
\end{equation}

\noindent
(note that the components of $g$ are not to be confused with our reduced spinor components), we can consider the right invariant 1-forms $\theta = d g \cdot g^{-1}$, and together with $\det(g) = 1$, we have
\begin{equation}
\theta = \begin{pmatrix} \bar{z}^{\bar{0}} dz^0 - \bar{z}^{\bar{1}} dz^1 & z^0 dz^1 - z^1 d z^0 \\ \bar{z}^{\bar{0}} d\bar{z}^{\bar{1}} - \bar{z}^{\bar{1}} d\bar{z}^{\bar{0}} & \bar{z}^{\bar{1}} dz^1 - \bar{z}^{\bar{0}} dz^0 \end{pmatrix}\,.
\end{equation}

\noindent
Using the basis Eq.(\ref{eq:reduc8.5}), we can expand $\theta = a \tau_1 + b \tau_2 + c \tau_3$ with
\begin{align}
a &= 2 i (\bar{z}^{\bar{0}} dz^0 - \bar{z}^{\bar{1}} dz^1)\,,\\[0.3\baselineskip]
b &= 2 \R(z^0 dz^1 - z^1 d z^0)\,,\\[0.3\baselineskip]
c &= -2 \I(z^0 dz^1 - z^1 d z^0)\,.
\end{align}

\noindent
The coefficients $b$ and $c$ are obviously real. To show that $a$ is real as well, use again $\det(g) = 1$. To obtain the symplectic structure on the different coadjoint orbits we have to consider certain representatives of those orbits, for example, $f_1=(s,0,0)$, $f_2=(0,s,0)$, or $f_3=(0,0,s)$. We get, for example,
\begin{equation}
\theta_{f_1} = 2is (\bar{z}^{\bar{0}} dz^0 - \bar{z}^{\bar{1}} dz^1)
\end{equation}

\noindent
which leads to
\begin{equation}
\omega_1 = -d\theta_{f_1} = 2i s (dz^0 \wedge d\bar{z}^{\bar{0}} - dz^1 \wedge d\bar{z}^{\bar{1}})\,.
\end{equation}

\noindent
This symplectic 2-forms induces the following Poisson bracket for functions $f,g$ on the coadjoint orbit of $f_1$,
\begin{align}
&\pb{f}{g}_1 =\\[0.3\baselineskip]
&2is \left(\frac{\partial f}{\partial z^0} \frac{\partial g}{\partial \bar{z}^{\bar{0}}} - \frac{\partial f}{\partial \bar{z}^{\bar{0}}} \frac{\partial g}{\partial z^0} - \frac{\partial f}{\partial z^1} \frac{\partial g}{\partial \bar{z}^{\bar{1}}} + \frac{\partial f}{\partial \bar{z}^{\bar{1}}} \frac{\partial g}{\partial z^1}\right)\,.\notag
\end{align}

\noindent
Hence, for $s = \frac{1}{2}$, we get the Poisson structure
\begin{equation}
\pb{z^A}{\bar{z}^{\bar{B}}}_1 = i (\sigma_3)^{A \bar{B}}\,,
\label{eq:coadpoisson1}
\end{equation}

\noindent
which is exactly Eq.(\ref{eq:dirac11}). Note, that we can choose different values for $s$, even negative ones. Using the coadjoint representation of a $g \in \s11$, we can build a representation of $\as11$ using those spinors and the Poisson brackets. Consider, for example,
\begin{align}
J_3 &\equiv \abs{z^0}^2 + \abs{z^1}^2\,,\notag\\[0.3\baselineskip]
K_1 &\equiv 2 \I(\bar{z}^{\bar{0}} z^1)\,,\\[0.3\baselineskip]
K_2 &\equiv 2 \R(\bar{z}^{\bar{0}} z^1)\,.\notag
\end{align}

\noindent
Together with the Poisson bracket Eq.(\ref{eq:coadpoisson1}), one shows that this gives indeed a (vector) representation of $\as11$ with
\begin{equation}
\pb{J_3}{K_1}_1 = 2 K_2\,,
\end{equation}
\begin{equation}
\pb{J_3}{K_2}_1 = -2 K_1\,,
\end{equation}
\begin{equation}
\pb{K_1}{K_2}_1 = -2 J_3\,.
\end{equation}

\noindent
Using the other coadjoint orbits $f_2$ or $f_3$, one can similarly construct different representations of $\as11$.

\section{Timelike twisted geometries}
\label{sec:fullgraph}
In this section we show that the twisted geometries parametrization of the phase space variables $(g,\Pi) \in \bsl2c$ in terms of normal vectors and angles is still valid in our case. We define for $\norm{\omega} / \norm{\lambda} \neq 0$
\begin{equation}
\Xi \equiv 2 \ln\left(\frac{\norm{\omega}}{\norm{\lambda}}\right)\,,
\label{eq:twist1}
\end{equation}

\noindent
where we have as usual $\norm{\omega}^2 = -(\sigma_3)_{A \bar{B}} \omega^A \bar{\omega}^{\bar{B}}$. Using the original Poisson bracket on twistor space (and not the reduced Dirac bracket), this new variable satisfies
\begin{equation}
\pb{\pi \omega}{\Xi} = 1 \quad , \quad \pb{\bar{\pi} \bar{\omega}}{\Xi} = 1\,,
\label{eq:twist2}
\end{equation}

\noindent
which, in turn, gives $\pb{\R(\pi \omega)}{\Xi} = 1$. If we consider furthermore the two normals that are associated with the source and target node of some link, respectively, we want to calculate the scalar product between those two normals. If we take the normal on the source node to be given by
\begin{equation}
n^{A \bar{B}}_s = \frac{i}{\sqrt{2}} (\sigma_3)^{A \bar{B}}
\label{eq:twist3}
\end{equation}

\noindent
and the one on the target node to be parallel transported with the $\sl2c$ holonomy from Eq.(\ref{eq:intro106}), i.e.,
\begin{equation}
n^{A \bar{B}}_t = \tensor[]{g}{^A_C} \tensor[]{\bar{g}}{^{\bar{B}}_{\bar{D}}} n^{C \bar{D}}_s\,,
\label{eq:twist4}
\end{equation}

\noindent
one finds that on the simplicity constraint $F_1 = 0$ and the area matching constraint $C=0$ we have
\begin{align}
\bk{n_t}{n_s} &= -\frac{1}{2} \left(\frac{\norm{\lambda}^2}{\norm{\omega}^2} + \frac{\norm{\omega}^2}{\norm{\lambda}^2}\right)\notag\\[0.3\baselineskip]
&= - \cosh(\Xi)\,.
\label{eq:twist5}
\end{align}

\noindent
Hence, the angle $\Xi$, as in the standard time-gauge case \cite{twisted4}, corresponds to the extrinsic curvature of the embedding of our 2+1 hypersurface in spacetime. The difference, however, is that it corresponds now to a boost angle on the one-sheeted hyperboloid and not the two-sheeted hyperboloid as in the standard time-gauge case. This makes sense because in both cases $F_1$ generates noncompact gauge orbits for real Barbero-Immirzi parameter, as can be seen from Eqs. (\ref{eq:class21}) - Eq.(\ref{eq:class24}). This result holds furthermore for $G_1 = 0$ as well. Now, is this angle still the conjugate variable to the area? We can use the Plebanski 2-form to define our area (squared) as $\mathcal{A}^2 \equiv \frac{1}{2} \Sigma^2$ and as defined in Eq.(\ref{eq:tscnew100}):
\begin{equation}
\mathcal{A}^2 = \frac{\gamma^2}{4} \, \R\left(\frac{(\pi \omega)^2}{(\gamma + i)^2}\right)\,.
\label{eq:areadef1}
\end{equation}

\noindent
Now, we can consider the Poisson bracket between the area $\mathcal{A}$ and the angle $\Xi$ to obtain
\begin{equation}
\pb{\mathcal{A}}{\Xi} = \frac{\gamma^2}{2 (1+\gamma^2)}\,.
\label{eq:twist8}
\end{equation}

\noindent
Thus, we see that, indeed, $\Xi$ and the area $\mathcal{A}$ are conjugate variables. Now, consider again the holonomy Eq.(\ref{eq:intro106}):
\begin{equation}
\tensor[]{g}{^A_B} = \frac{\lambda^A \pi_B + \sigma^A \omega_B}{\sqrt{\pi \omega} \sqrt{\lambda \sigma}}\,.
\label{eq:twist9}
\end{equation}

\noindent
Following Ref. \cite{twisted4}, we can write it as a product of two matrices,
\begin{equation}
g = \mathrm{m}(\sigma,\lambda) \, \mathrm{m}(-\pi,\omega)^{-1}
\label{eq:twist10}
\end{equation}

\noindent
with
\begin{equation}
\mathrm{m}(\sigma,\lambda) = \frac{i}{\sqrt{\lambda \sigma}} \begin{pmatrix} \sigma^0 & \lambda^0 \\ \sigma^1 & \lambda^1 \end{pmatrix} \quad , \quad \det(\mathrm{m}) = 1
\label{eq:twist11}
\end{equation}

\noindent
and
\begin{align}
\mathrm{m}(-\pi,\omega) &= \frac{i}{\sqrt{-\omega \pi}} \begin{pmatrix} -\pi^0 & \omega^0 \\ -\pi^1 & \omega^1 \end{pmatrix} \quad , \quad \det(\mathrm{m}) = 1\,,\notag\\[0.3\baselineskip]
&= \frac{i}{\sqrt{\pi \omega}} \begin{pmatrix} -\pi_1 & \omega_1 \\ \pi_0 & -\omega_0 \end{pmatrix}
\label{eq:twist12}
\end{align}

\noindent
and we have
\begin{equation}
\mathrm{m}(-\pi,\omega)^{-1} = \frac{-i}{\sqrt{\pi \omega}} \begin{pmatrix} \omega_0 & \omega_1 \\ \pi_0 & \pi_1 \end{pmatrix}\,.
\label{eq:twist13}
\end{equation}

\noindent
In comparison with Ref. \cite{twisted4}, we have introduced the extra $i$ factor in order to have $\det(\mathrm{m}) = 1$ and not $\det(\mathrm{m}) = -1$. This has the advantage that these matrices $\mathrm{m}$ are elements of $\sl2c$ and not just of the general linear group (which is not semisimple). Since $g, \mathrm{m} \in \sl2c$, we can use the Iwasawa decomposition (for semisimple Lie groups) for both $\mathrm{m}$ and express the holonomy $g$ in terms of a $\su2$ matrix, an upper-triangular matrix, and a diagonal boost. However, since we are interested in a reduction of $\sl2c$ down to $\s11$, we propose an Iwasawa-like decomposition of the holonomy that includes $\s11$ as follows. We write for an arbitrary element $g \in \sl2c$
\begin{align}
g &= \begin{pmatrix} \alpha & \beta \\ \gamma & \delta \end{pmatrix}\notag\\[0.3\baselineskip]
&= \begin{pmatrix} e & f \\ \bar{f} & \bar{e} \end{pmatrix} \begin{pmatrix} 1 & n \\ 0 & 1 \end{pmatrix} \begin{pmatrix} i & 0 \\ 0 & -i \end{pmatrix}^{\epsilon} \begin{pmatrix} t & 0 \\ 0 & t^{-1} \end{pmatrix}\,,
\label{eq:twist14}
\end{align} 

\noindent
where the first factor is a matrix in $\s11$, because $\abs{e}^2 - \abs{f}^2 = 1$, which follows from $\det(g) = 1$. We find ($t \in \mathbb{R}_{>0}$)
\begin{equation}
t \equiv \begin{cases}
\: \: \: \sqrt{\abs{\alpha}^2 - \abs{\gamma}^2} \: , \quad \text{for} \quad \abs{\alpha}^2 > \abs{\gamma}^2\\
\\
\: \: \: \sqrt{\abs{\gamma}^2 - \abs{\alpha}^2} \: , \quad \text{for} \quad \abs{\alpha}^2 < \abs{\gamma}^2
\end{cases}\,,
\label{eq:twist15}
\end{equation}

\noindent
where in the first case we have $\epsilon = 0$ and in the second case we have $\epsilon = 1$. Without this discrete variable $\epsilon$ we would not be able to cover the whole $\sl2c$ manifold away from the identity in the above manner. With this definition for $t$, we find further ($e,f \in \mathbb{C}$)
\begin{equation}
e = \frac{\alpha}{t} \qquad , \qquad f=\frac{\bar{\gamma}}{t}
\label{eq:twist16}
\end{equation}

\noindent
and
\begin{equation}
n = - \frac{\bar{\gamma}}{\alpha} + \frac{\beta}{\alpha} \, t^2 = - \frac{\bar{\alpha}}{\gamma} + \frac{\delta}{\gamma} \, t^2 \in \mathbb{C}\,.
\label{eq:twist17}
\end{equation}

\noindent
If we express now $\mathrm{m}(\sigma,\lambda)$ and $\mathrm{m}(-\pi,\omega)$ in this parametrization, we see that $g \in \s11$ iff
\begin{equation}
t = \utilde{t} \qquad \text{and} \qquad n = - \utilde{n}\,,
\label{eq:twist18}
\end{equation}

\noindent
which is one real and one complex constraint. Note that the decomposition in Eq.(\ref{eq:twist14}) is different from the one used in Ref. \cite{twisted4}, not only because we consider an $\s11$ element in the first factor, but also because we consider a pure boost for the last matrix. In Ref. \cite{twisted4}, the authors use a combination of boost and rotation. If we chose a different expression for our $\s11$ element, we could try to obtain a similar decomposition, which, of course, would also give a different expression for the simplicity constraints again. It would be interesting to see how the angle $\Xi$, defined in Eq.(\ref{eq:twist1}), would enter such a decomposition, and one should furthermore obtain a reparametrization for the fluxes as well, but we will leave this for future investigations. Before we consider the quantization of this model in Sec. \ref{sec:quant} we will first investigate a general graph, instead of a single link, and consider the reduction by the closure constraint.

\subsection{Closure constraint}
\label{sec:closureconstraint}
We consider now a general graph $\Gamma$ with $L$ links and $N$ nodes. At each of these nodes, we aim to impose local gauge invariance under $\s11$ transformations via the so-called closure or Gauss constraint. We have shown that the symplectic reduction for a single link phase space by area matching and simplicity constraints gives $\mathbb{T}^2_{\ast} \sslash C \sslash F_{1,2} \simeq \bs11$. Hence, for a graph with $L$ links and $N$ nodes, we have
\begin{equation}
\mathbb{T}^{2L}_{\ast} \sslash C_l \sslash F_{l,1,2} \simeq \bs11^L\,.
\end{equation}

\noindent
Now, we want to further investigate what happens if we take the Gauss constraint into account. This constraint (in its covariant form) is given by
\begin{equation}
\mathcal{G}^{IJ}_n \equiv \sum_{l_i \in n}{B^{IJ}_{l_i}} = 0
\end{equation}

\noindent
for each node $n$ of the graph and imposes local $\sl2c$ gauge invariance. On the unconstrained level we can express $\mathcal{G}^{IJ}_n$ in terms of the self-dual components $\Pi^{AC}_l \epsilon^{\bar{B} \bar{D}}$ as
\begin{equation}
\tilde{\mathcal{G}}^{AC}_n \equiv \sum_{l_i \in n}{\Pi^{AC}_{l_i}} = 0\,,
\end{equation}

\noindent
which is enough to guarantee $\mathcal{G}^{IJ}_n = 0$. It should be clear that this constraint interacts with each link always with just one term. Hence, it is easy to show that
\begin{equation}
\pb{\tilde{\mathcal{G}}^{AC}_n}{C_{l_j}} = 0 = \pb{\tilde{\mathcal{G}}^{AC}_n}{\mathring{F}_{1,l_j}}\,.
\end{equation}

\noindent
However, not surprisingly, with the second-class constraints $F_{2,l}$, we find that
\begin{align}
&\pb{\tilde{\mathcal{G}}^{AC}_n}{F_{2,l_j}} = \sum_{l_i \in n}{\pb{\Pi^{AC}_{l_i}}{F_{2,l_j}}}\\[0.3\baselineskip]
&\qquad = - \frac{1}{4} \left(\pi^A_j n^{C \bar{B}} \bar{\omega}_{j,\bar{B}} + \pi^C_j n^{A \bar{B}} \bar{\omega}_{j,\bar{B}}\right) \neq 0\,,\notag
\end{align}

\noindent
which should be obvious because the $F_{2,l}$ are just invariant under the little group $\s11$. On the other hand, if we consider again the master constraint, we obtain 
\begin{align}
\pb{\tilde{\mathcal{G}}^{AC}_n}{\textbf{M}_l} &= \pb{\tilde{\mathcal{G}}^{AC}_n}{\bar{F}_{2,l_j} F_{2,l_j}}\notag\\[0.3\baselineskip]
&= \bar{F}_{2,l_j} \pb{\tilde{\mathcal{G}}^{AC}_n}{F_{2,l_j}} + \pb{\tilde{\mathcal{G}}^{AC}_n}{\bar{F}_{2,l_j}} F_{2,l_j}\notag\\[0.3\baselineskip]
&\approx 0\,.
\end{align}

\noindent
This means that when we consider the master constraints $\textbf{M}_l$, (together with the $C_l$ and $F_{1,l}$) we have a system of only first-class constraints. Furthermore, if we impose first area matching and covariant closure constraints, which leads to $\sl2c$ BF theory, we can in principle consider $\sl2c$ intertwiners, which are then further reduced to intertwiners of the little group, i.e., $\su2$ or $\s11$, upon the imposition of the remaining simplicity constraints. Now, if we consider the reduced phase space $\mathbb{C}^2 \times \mathbb{C}^2$, where we solved the simplicity constraints already, then we are left with the reduced area matching constraint $C_{\text{red}}$ and a reduced version of the closure constraint that generates local gauge transformations of the little group. In particular, we can write for the reduced closure constraint
\begin{equation}
\mathring{G}^i_{n} = \sum_{l_i \in n}{\pi^{i}_{l_i}} = 0\,.
\end{equation}

\noindent
Since the $\pi^i$ are gauge invariant with respect to the reduced area matching constraint, we have
\begin{equation}
\pb{\mathring{G}^i_{n}}{C_{\text{red},l_j}} = 0
\end{equation}

\noindent
and hence on $(\mathbb{C}^2 \times \mathbb{C}^2)^L$, we can consider $L$ reduced area matching constraints and $N$ reduced closure constraints. All are first-class and hence we get that the dimension of the graph Hilbert space is $8L - 2L - 3 \times 2 \times N = 6(L-N)$, exactly as in the timelike case. (Note that there are three closure constraints per node, one for each component $i$.)

In the quantum theory, the solution space of the (reduced) closure constraint leads to $\s11$ spin networks where the nodes are decorated with $\s11$ intertwiners. We refer the reader to Refs. \cite{sellaroli1,sellaroli2,sellaroli3} for details on those intertwiners, which require more care than their $\su2$ analogs.

\section{Quantization and timelike spin networks}
\label{sec:quant}
Our starting point for the quantization, following Refs. \cite{twisted4}, \cite{twistphasespace}, and \cite{nulltwisted}, are quantum twistor networks, which are graphs labeled with 2-twistor space $\mathbb{T}^2_{\ast}$ on each link. This space $\mathbb{T}_{\ast}$, one for each half-link, can easily be quantized by promoting the spinorial components of the twistors to operators and their Poisson brackets to the corresponding commutators in a Schr\"{o}dinger representation. This will provide us with our unconstrained Hilbert space on which we then impose the quantized simplicity constraints, (reduced) area matching constraint, and closure constraints (in this order). For each link, we consider the auxiliary Hilbert space of homogeneous functions of degree $(a,b)$. Hence, we consider $f : \mathbb{C}^2 \longrightarrow \mathbb{C}$ such that $\forall \lambda \in \mathbb{C}_{\ast}$,
\begin{equation}
f(\lambda \omega^A) = \lambda^a \bar{\lambda}^b f(\omega^A)\,.
\label{eq:quant1}
\end{equation}

\noindent
These functions are essentially functions on $\mathbb{CP}^1$. To deal with single valued functions, we have to require that $a-b$ must be an integer. Note, furthermore, that these functions are not assumed to be holomorphic or antiholomorphic, since they are general polynomials in the spinor components as well as their complex conjugates. In certain cases, however, they can be reduced to give holomorphic representations. Together with
\begin{equation}
(g \triangleright f)(\omega^A) = f(g^{-1} \triangleright \omega^A)\,,
\label{eq:quant2}
\end{equation}

\noindent
this provides, for certain values of the numbers $(a,b)$,  a unitary and irreducible representation for $\sl2c$ \cite{ruhl}. The $\sl2c$-invariant measure on this space of functions is given by
\begin{equation}
d\Omega(\omega^A) = \frac{i}{2} (\omega^0 d \omega^1 - \omega^1 d \omega^0) \wedge (\bar{\omega}^{\bar{0}} d \bar{\omega}^{\bar{1}} - \bar{\omega}^{\bar{1}} d \bar{\omega}^{\bar{0}})\,.
\label{eq:quant3}
\end{equation}

\noindent
Under rescaling, it transforms as $d\Omega(\lambda \omega^A) = \abs{\lambda}^4 \, d\Omega(\omega^A)$ so that the $\sl2c$ and scaling-invariant scalar product is given by
\begin{equation}
\bk{f_1}{f_2} = \frac{i}{2} \int_{\mathbb{CP}^1}{d\Omega(\omega^A) \, \bar{f}_1(\omega^A) f_2(\omega^A)}\,.
\label{eq:quant4}
\end{equation}

\noindent
This representation belongs to the principal series of $\sl2c$. With $n \in \mathbb{Z}/2$ and $p \in \mathbb{R}$, it is unitary, and we denote the corresponding Hilbert space of those functions by $\mathcal{H}^{(n,p)}$. The numbers $(a,b)$ and $(n,p)$ are related by
\begin{equation}
a = -n -1 +ip \qquad \text{and} \qquad b = n - 1 + ip\,. 
\label{eq:quant5}
\end{equation}

\noindent
Since the representations $(n,p)$ and $(-n,-p)$ are unitarily equivalent, we restrict those labels to be $n \in \mathbb{N}_0/2$ and $p \in \mathbb{R}$. The labels $(n,p)$ are related to the eigenvalues of the $\asl2c$ Casimirs $C_1 = \vec{L}^2 - \vec{K}^2$ and $C_2 = -2 \vec{L} \cdot \vec{K}$ as follows:
\begin{align}
&\hat{C}_1 \, \triangleright \, f^{(n,p)} = (n^2 - p^2 - 1) f^{(n,p)}\,,\label{eq:quant6}\\[0.3\baselineskip]
&\hat{C}_2 \, \triangleright \, f^{(n,p)} = -2 n p \, f^{(n,p)}\,.\label{eq:quant6.5}
\end{align}

\noindent
Note that under the change $(n,p) \mapsto (-n,-p)$ the Casimir $C_1$ stays the same, whereas $C_2$ changes its sign. If we consider the half-link phase space $\mathbb{T}_{\ast}$ with $Z^{\alpha} = (\omega^A, i \bar{\pi}_{\bar{B}})$ and $\pi \omega = \epsilon_{AB} \pi^A \omega^B \neq 0$, the Poisson structure of which is given by
\begin{equation}
\pb{\pi_A}{\omega^B} = \delta^B_A \quad , \quad \pb{\bar{\pi}_{\bar{A}}}{\bar{\omega}^{\bar{B}}} = \delta^{\bar{B}}_{\bar{A}}\,,
\label{eq:quant7}
\end{equation}

\noindent
and similarly for $W^{\alpha} = (\lambda^A, i \bar{\sigma}_{\bar{B}})$ with $\sigma \lambda \neq 0$, we use for the commutators
\begin{equation}
\comm{\hat{\pi}_{A}}{\hat{\omega}^{B}} = - i \hbar \, \delta^B_A \quad , \quad \comm{\hat{\bar{\pi}}_{\bar{A}}}{\hat{\bar{\omega}}^{\bar{B}}} = - i \hbar \, \delta^{\bar{B}}_{\bar{A}}
\label{eq:quant8}
\end{equation}

\noindent
the following Schr\"{o}dinger representation:
\begin{align}
&\hat{\omega}^B f(\omega^A) = \omega^B f(\omega^A)\,,\label{eq:quant9}\\[0.3\baselineskip]
&\hat{\pi}_B f(\omega^A) = - i \hbar \, \frac{\partial}{\partial \omega^B} f(\omega^A)\,.\label{eq:quant9.5}
\end{align}

\noindent
The homogeneous functions are furthermore interesting because they diagonalize the Euler dilatation operator $\omega^A \partial_A$,
\begin{align}
\omega^A \frac{\partial}{\partial \omega^A} f^{(a,b)}(\omega^A) &= a \, f^{(a,b)}(\omega^A)\,,\label{eq:quant10}\\[0.3\baselineskip]
\bar{\omega}^{\bar{A}} \frac{\partial}{\partial \bar{\omega}^{\bar{A}}} f^{(a,b)}(\omega^A) &= b \, f^{(a,b)}(\omega^A)\,,\label{eq:quant10.5}
\end{align}

\noindent
which holds for all homogeneous functions. The Hilbert space for each single link is now given by the homogeneous functions of the form
\begin{equation}
f^{(a,b)}(\omega^A,\lambda^B) \equiv f^{(a_s,b_s)}(\omega^A) \otimes f^{(a_t,b_t)}(\lambda^A)\,,
\label{eq:quant11}
\end{equation}

\noindent
where the subscripts $s$ and $t$ stand for the source and target half-links. It is easy to see that these are now homogeneous functions of degree $(a,b) = (a_s + a_t,b_s+b_t)$. Recall that the complex area matching constraint Eq.(\ref{eq:intro105}) was given by $C = \pi \omega - \lambda \sigma = 0$. We can use Eq.(\ref{eq:quant10}) to impose $\hat{C} = 0$ as follows. We can write $\: \pi \omega = \pi_A \omega^A = \frac{1}{2} (\pi \omega + \pi \omega) = \frac{1}{2} (\pi \omega - \omega \pi)$. This gives us a normal ordering for $\widehat{\pi \omega}$,
\begin{align}
\widehat{\pi \omega} &= \frac{\hbar}{2i} \left[\frac{\partial}{\partial \omega^A} \omega^A - \omega_A \frac{\partial}{\partial \omega_A}\right]\notag\\[0.3\baselineskip]
&= \frac{\hbar}{2i} \left[\omega^A \frac{\partial}{\partial \omega^A} + \frac{\partial}{\partial \omega^A} \omega^A\right]\,,
\label{eq:quant12}
\end{align}

\noindent
where we have used that switching the position of spinorial indices gives a minus sign in the second equality. Analogously one obtains for the complex conjugate contribution
\begin{equation}
\widehat{\bar{\pi} \bar{\omega}} = \frac{\hbar}{2i} \left[\bar{\omega}^{\bar{A}} \frac{\partial}{\partial \bar{\omega}^{\bar{A}}} + \frac{\partial}{\partial \bar{\omega}^{\bar{A}}} \bar{\omega}^{\bar{A}}\right]
\label{eq:quant13}
\end{equation}

\noindent
and the corresponding expressions in terms of $(\sigma,\lambda)$ variables. Using now the commutation relations and Eq.(\ref{eq:quant10}), we can show that for a homogeneous function with degree $(a,b)$ we have
\begin{align}
\widehat{\pi \omega} \, f^{(a,b)} &= \frac{\hbar}{2 i} \left[\omega^A \frac{\partial}{\partial \omega^A} + \frac{\partial}{\partial \omega^A} \omega^A\right] f^{(a,b)}\notag\\[0.3\baselineskip]
&= \frac{\hbar}{2 i} \left[\omega^A \frac{\partial}{\partial \omega^A} + 2 + \omega^A \frac{\partial}{\partial \omega^A}\right] f^{(a,b)}\notag\\[0.3\baselineskip]
&= \frac{\hbar}{i} \left[a + 1\right] f^{(a,b)}
\label{eq:quant14}
\end{align}

\noindent
and similarly
\begin{equation}
\widehat{\bar{\pi} \bar{\omega}} \, f^{(a,b)} = \frac{\hbar}{i} \left[b + 1\right] f^{(a,b)}\,.
\label{eq:quant15}
\end{equation}

\noindent
The action of the area-matching constraint becomes
\begin{align}
&\hat{C} \, \triangleright f^{(a,b)}(\omega^A,\lambda^B) = \hat{C} \, \triangleright \left(f^{(a_s,b_s)}(\omega^A) \otimes f^{(a_t,b_t)}(\lambda^A)\right)\notag\\[0.3\baselineskip]
&= \left(\hat{C} \otimes 1 + 1 \otimes \hat{C}\right) \left(f^{(a_s,b_s)}(\omega^A) \otimes f^{(a_t,b_t)}(\lambda^A)\right)\notag\\[0.3\baselineskip]
&= \left(\widehat{\pi \omega} \triangleright f^{(a_s,b_s)}(\omega^A)\right) \otimes f^{(a_t,b_t)}(\lambda^A)\notag\\[0.3\baselineskip]
&\qquad \qquad - f^{(a_s,b_s)}(\omega^A) \otimes \left(\widehat{\lambda \sigma} \triangleright f^{(a_t,b_t)}(\lambda^A)\right)\notag\\[0.3\baselineskip]
&= \frac{\hbar}{i} \left[a_s + a_t + 2\right] \left(f^{(a_s,b_s)}(\omega^A) \otimes f^{(a_t,b_t)}(\lambda^A)\right)
\label{eq:quant16}
\end{align}

\noindent
and analogously the complex conjugate area-matching constraint gives
\begin{equation}
\hat{\bar{C}} \, \triangleright f^{(a,b)}(\omega^A,\lambda^B) = \frac{\hbar}{i} \left[b_s + b_t + 2\right] f^{(a,b)}(\omega^A,\lambda^B)\,.
\label{eq:quant17}
\end{equation}

\noindent
Using Eq.(\ref{eq:quant5}), one finds that $a_s + a_t + 2 = -(n_s + n_t) + i (p_s + p_t)$, and hence both constraints are solved by $n_t = -n_s$ and $p_t = -p_s$. Since we want to work with $n_i \in \frac{\mathbb{N}_0}{2}$, we have to consider on the source link states with $(n_s,p_s)$ and on the target link states with $(-n_s,-p_s)$, which are states from two different (but unitarily equivalent) Hilbert spaces.

Before we investigate the imposition of the simplicity constraints in the next sections, we recall that the so-called canonical basis for $\mathcal{H}^{(n,p)}$, which stems from an induced representation using the $\su2$ subgroup of $\sl2c$, is used in the quantization of the EPRL model using the time gauge. This is possible because we can further diagonalize $\vec{L}^2$ and $L^3$ besides the two $\asl2c$ Casimirs, which gives the states $\ket{(n,p);j,m}$, where $j \in \mathbb{N}_0/2$ denotes the spin and $m \in \{-j,-j+1,\cdots,j\}$ denotes its magnetic number. In particular, this leads to a decomposition of $\mathcal{H}^{(n,p)}$ as
\begin{equation}
\mathcal{H}^{(n,p)} \simeq \bigoplus_{n \leq j} \mathcal{H}^{(j)}\,,
\label{eq:quant18}
\end{equation}

\noindent
where $\mathcal{H}^{(j)}$ denotes the standard $(2j+1)$-dimensional unitary and irreducible representation space of $\su2$. Since the stabilizing subgroup for our spacelike normal vector $N^I =(0,0,0,1)^t$ is given by $\s11$, it is more suitable to employ a decomposition in terms of a $\s11$ basis. This was also used in Refs. \cite{LorentzianSFM1} and \cite{LorentzianSFM2}. For that reason, we briefly review some representation theory of $\s11$ in the following section.

\subsection{Representations of \texorpdfstring{$\s11$}{SU(1,1)}}
\label{sec:irreps}
The $\sl2c$ representations from above provide, of course, representations for the subgroup $\s11$ as well. They are, however, not irreducible. But similarly to Eq.(\ref{eq:quant18}), they can be decomposed into $\s11$ irreducible representations. To fix our conventions, we consider here the unitary and irreducible representations of $\s11$ belonging to the principal series. The early works on the representation theory of the three-dimensional Lorentz group are Ref. \cite{bargmann} or the book Ref. \cite{gelfand}. The Plancherel decomposition was investigated, for example, in Ref. \cite{harish2}, and for a newer account, see Ref. \cite{ruhl}. The Clebsch-Gordan problem for $\s11$ was investigated in Refs. \cite{decomp1,decomp2,decomp3,decomp4}. Note, that in this work we have so far used the mathematical convention for the rotation and boost generators, i.e., $L^{\dagger}_i = -L_i$ and $K^{\dagger}_i = K_i$. In Refs. \cite{LorentzianSFM1} and \cite{LorentzianSFM2} or \cite{sellaroli1,sellaroli2,sellaroli3}, for example, the authors use the physical convention where the Hermiticity property is reversed. This will not be an obstacle in what follows, since the simplicity constraints are invariant under this choice. This can easily be seen from Eq.(\ref{eq:intro114}), where one can simply define the $\Pi^i$ with an additional factor of $\pm i$ and this would not change the form of the master constraint, as can be seen from Eqs. (\ref{eq:tsc16}) - (\ref{eq:tsc19}). For the covariant simplicity constraints $F_1$ and $G_1$, this convention is irrelevant as well, since for them we do not use the generators $L_i$ and $K_i$ explicitly. Now, with this in mind, we can consider the physical convention, where $L_3$ is Hermitian and hence can be diagonalized with a real eigenvalue. Furthermore, we look for states that diagonalize the $\as11$ Casimir $Q_{\as11} = (L^3)^2 - (K^1)^2 - (K^2)^2$. We denote those eigenstates of the two $\asl2c$ Casimirs $C_1$ and $C_2$ as well as $Q_{\as11}$ and $L^3$ by $f^{(n,p)}_{j,m} = \ket{(n,p);j,m} \in \mathcal{H}^{(n,p)}$. The eigenvalues of the $\asl2c$ Casimirs are given by Eqs. (\ref{eq:quant6}) and (\ref{eq:quant6.5}), and we have furthermore
\begin{align}
Q_{\as11} \, \triangleright \, f^{(n,p)}_{j,m} &= \pm j(j+1) \, f^{(n,p)}_{j,m}\,,\label{eq:repr1}\\[0.3\baselineskip]
L^3 \, \triangleright \, f^{(n,p)}_{j,m} &= m \, f^{(n,p)}_{j,m}\,.\label{eq:repr1.5}
\end{align}

\noindent
The action of $Q_{\as11}$ with a plus is the convention as used, for example, in Refs. \cite{LorentzianSFM1} and \cite{LorentzianSFM2}, whereas in Ref. \cite{sellaroli3} the authors use the additional minus sign in front of $j(j+1)$. We will see that this sign plays a role for our final result. We will find that the solutions to the master constraint with the discrete states on both half-links do not give us the full reduced Hilbert space necessary to decompose all functions on $\s11$ in a spin network basis\footnote{Rather, one would obtain only the discrete states with integer spin and the continuous states with even parity.}. Hence, we are eventually forced to work with the convention from Ref. \cite{sellaroli3}, i.e., with eigenvalues $-j(j+1)$. Furthermore, let us point out that if we compare our approach with the coherent state approach used in Refs. \cite{LorentzianSFM1} and \cite{LorentzianSFM2}, where it was stated that it is necessary to diagonalize a noncompact generator $K^1$ or $K^2$ instead of $L^3$, in order to be able to describe timelike faces, we do not find this to be necessary, which makes our considerations more comprehensible.

For $\s11$, we have the following unitary irreducible representations (that appear in the Plancherel decomposition), which are all infinite dimensional, since $\s11$ is noncompact. First, we have the discrete series $\mathcal{D}^{\pm}_k$ where $j = -k$ with $k \in \frac{\mathbb{N}}{2}$. For $\mathcal{D}^{+}_k$, we have $m \in \{k, k+1, k+2, \cdots \}$, and for $\mathcal{D}^{-}_k$, we have $m \in \{-k, -k-1, -k-2, \cdots \}$. The state with $j = -1/2$ is somewhat special in that it is not normalizable and hence does not appear in the Plancherel decomposition. We see that using the plus convention in Eq.(\ref{eq:repr1}) and if we do not consider the state with $j=-1/2$  then we have for all other possible values of $j$ in the discrete series 
\begin{equation}
Q^d_{\s11} \in \{0, \frac{3}{4}, 2, \frac{15}{4}, \cdots\} \geq 0\,.
\label{eq:repr2}
\end{equation}

\noindent
Second, we have the continuous series $\mathcal{C}^{\varepsilon}_s$ with $j = - \frac{1}{2} + i s$ and $\varepsilon \in \{0,\frac{1}{2}\}$. For $\varepsilon = 0$ (even functions), we have $s \geq 0$ and $m \in \{0, \pm 1, \pm 2, \cdots \}$, and for $\varepsilon = \frac{1}{2}$ (odd functions), we have $s > 0$ and $m \in \{\pm \frac{1}{2}, \pm \frac{3}{2}, \pm \frac{5}{2}, \cdots \}$. Hence, using again the plus convention in Eq.(\ref{eq:repr1}), we have for all states from $\mathcal{C}^{\varepsilon}_s$
\begin{equation}
Q^c_{\s11} = j(j+1) = -s^2 - \frac{1}{4} < 0\,.
\label{eq:repr3}
\end{equation}

\noindent
In what follows, we will first use this convention and only later change to the opposite case. We explicitly include the full analysis in order to pinpoint exactly where the problem with this convention lies. We just mention that the analog of Eq.(\ref{eq:quant18}) reads in this noncanonical basis \cite{ruhl,unitaryirrep}
\begin{align}
\mathcal{H}^{(n,p)} &\simeq \left(\bigoplus^{n}_{k>1/2} \mathcal{D}^+_k \oplus \int^{\infty \oplus}_0{ds \, \mathcal{C}^{\varepsilon}_s}\right)\notag\\[0.3\baselineskip]
&\qquad \oplus \left(\bigoplus^{n}_{k>1/2} \mathcal{D}^-_k \oplus \int^{\infty \oplus}_0{ds \, \mathcal{C}^{\varepsilon}_s}\right)\,,
\label{eq:qu11}
\end{align}

\noindent
where the sum over the discrete states ranges over values for which $k-n$ is an integer and similarly $\varepsilon$ is determined by the condition that $\varepsilon - n$ is an integer. The Clebsch-Gordan decomposition for the coupling of those representations is given by \cite{decomp1,decomp2,decomp3,decomp4} 
\begin{equation}
\mathcal{D}^{\pm}_{k_1} \otimes \mathcal{D}^{\pm}_{k_2} = \bigoplus^{\infty}_{K = k_1 + k_2} \mathcal{D}^{\pm}_K\,,
\label{eq:repr4}
\end{equation}

\noindent
and
\begin{equation}
\mathcal{D}^{\pm}_{k_1} \otimes \mathcal{D}^{\mp}_{k_2} = \bigoplus^{k_1 - k_2}_{K=K_{\text{min}}} \mathcal{D}^{\pm}_K \: \oplus \bigoplus^{k_2 - k_1}_{K=K_{\text{min}}} \mathcal{D}^{\mp}_K \: \oplus \int^{\infty \oplus}_0{\mathcal{C}^{\varepsilon}_s \: ds}\,,
\label{eq:repr5}
\end{equation}

\noindent
where $K_{\text{min}} = 1$ and $\varepsilon = 0$ if $k_1 + k_2$ is an integer and $K_{\text{min}} = \frac{3}{2}$ and $\varepsilon = \frac{1}{2}$ otherwise. Furthermore, note that the discrete contributions vanish when the upper limits $k_1 - k_2$ or $k_2 - k_1$, respectively, are smaller than $1$; i.e, we must have $k_1 - k_2 \geq 1$ for the first sum and $k_2 - k_1 \geq 1$ for the second. The coupling of two continuous states gives
\begin{equation}
\mathcal{C}^{\varepsilon_1}_{s_1} \otimes \mathcal{C}^{\varepsilon_2}_{s_2} = \bigoplus^{\infty}_{K=K_{\text{min}}} \mathcal{D}^{+}_K \: \oplus \bigoplus^{\infty}_{K=K_{\text{min}}} \mathcal{D}^{-}_K \: \oplus 2 \int^{\infty \oplus}_0{\mathcal{C}^{\varepsilon}_s \: ds}\,,
\label{eq:repr6}
\end{equation}

\noindent
where $K_{\text{min}} = 1$ and $\varepsilon = 0$ if $\varepsilon_1 + \varepsilon_2 \in \mathbb{Z}$ and $K_{\text{min}} = \frac{3}{2}$ and $\varepsilon = \frac{1}{2}$ otherwise. The coupling of discrete states $k \in \frac{\mathbb{N}}{2}$ with continuous states $\varepsilon \in \{0,\frac{1}{2}\}$ and $0 < s < \infty$ gives
\begin{equation}
\mathcal{D}^{\pm}_{k} \otimes \mathcal{C}^{\varepsilon}_{s} = \bigoplus^{\infty}_{K=K_{\text{min}}} \mathcal{D}^{\pm}_K \: \oplus \int^{\infty \oplus}_0{\mathcal{C}^{\varepsilon'}_{s'} \: ds'}\,,
\label{eq:repr7}
\end{equation}

\noindent
where $K_{\text{min}} = 1$ and $\varepsilon' = 0$ if $k + \varepsilon$ is an integer and $K_{\text{min}} = \frac{3}{2}$ and $\varepsilon' = \frac{1}{2}$ otherwise. The Clebsch-Gordan coefficients for $\s11$ can be defined, and explicit formulas for their calculation can be found in Ref. \cite{klimyk1}. However, due to the noncompactness of $\s11$ and the different representation series, their explicit calculation is more complicated than in the $\su2$ case.

\subsection{Spacelike faces}
\label{sec:spacequant}
We consider now the imposition of the quantized simplicity constraints in the quantum theory. For the Lorentz-invariant part Eq.(\ref{eq:class3}), we use Eqs. (\ref{eq:quant14}) and (\ref{eq:quant15}) to obtain
\begin{align}
\hat{\mathring{F}}_1 \, f^{(a,b)} &= \left[(\gamma - i) \, \widehat{\pi \omega} - (\gamma + i) \, \widehat{\bar{\pi} \bar{\omega}}\right] f^{(a,b)}\notag\\[0.3\baselineskip]
&= \frac{\hbar}{i} \left[(\gamma - i) \, [a+1] - (\gamma + i) \, [b+1]\right] f^{(a,b)}\notag\\[0.3\baselineskip]
&= \frac{\hbar}{i} \left[\gamma [a-b] - i [a+b+2]\right] f^{(a,b)}\,.
\label{eq:qu17}
\end{align}

\noindent
In terms of the labels $(n,p)$, we have $a-b=-2n$ and $a+b+2=2ip$, and thus we get
\begin{align}
\hat{\mathring{F}}_1 \, f^{(a,b)} &= \frac{\hbar}{i} \left[-2 \gamma n +2p\right] f^{(a,b)} \stackrel{!}{=} 0\notag\\[0.3\baselineskip]
& \qquad \Leftrightarrow \quad p = \gamma n\,,
\label{eq:qu18}
\end{align}

\noindent
which is the well-known result from the EPRL model. Note that this provides a new way of describing spacelike faces in a nonstandard gauge and hence is interesting by itself. However, it is important to remember that our solution states $f^{(n,\gamma n)}_{j,m}$ (the master constraint not yet imposed) are not to be confused with the states one obtains with the standard time gauge. Those states are also denoted in the same way [or as $\ket{(n,\gamma n);j,m}$] but are very different states, because they diagonalize $\vec{L}^2$ and not $Q_{\as11}$. How to connect those states (when $j = -k$ for the discrete series) can be found in Refs. \cite{LorentzianSFM1} and \cite{LorentzianSFM2} or \cite{ruhl}.

\subsection{Timelike faces}
\label{sec:timequant}
For the dual constraint $\mathring{G}_1$, one obtains now similarly
\begin{align}
\hat{\mathring{G}}_1 \, f^{(a,b)} &= \left[(\gamma - i) \, \widehat{\pi \omega} + (\gamma + i) \, \widehat{\bar{\pi} \bar{\omega}}\right] f^{(a,b)}\notag\\[0.3\baselineskip]
&= \frac{\hbar}{i} \left[(\gamma - i) \, [a+1] + (\gamma + i) \, [b+1]\right] f^{(a,b)}\notag\\[0.3\baselineskip]
&= \frac{\hbar}{i} \left[\gamma [a+b+2] - i [a-b]\right] f^{(a,b)}
\label{eq:qu17.5}
\end{align}

\noindent
and again in terms of the labels $(n,p)$, we have $a+b+2=2ip$ and $a-b=-2n$, and thus we get
\begin{align}
\hat{\mathring{G}}_1 \, f^{(a,b)} &= 2 \hbar  \left[\gamma p + n\right] f^{(a,b)} \stackrel{!}{=} 0\notag\\[0.3\baselineskip]
& \qquad \Leftrightarrow \quad p = -\frac{n}{\gamma}\,.
\label{eq:qu18.5}
\end{align}

\noindent
This result was also found in Refs. \cite{LorentzianSFM1} and \cite{LorentzianSFM2}, and we will see in Sec. \ref{sec:area} that those states indeed can be associated to timelike faces\footnote{Note, furthermore, that this solution is also obtained from the first-class constraint mentioned in footnote 2.}. This is one of the main results of this paper. It not only confirms the solution found in Refs. \cite{LorentzianSFM1} and \cite{LorentzianSFM2} but, in fact, provides or more rigorous derivation, since it does not resort to some sort of large spin argument, which is typical for the coherent state approach to the imposition of the simplicity constraints. However, we will also see that we do not necessarily need those dual solutions in order to obtain timelike area spectra on the reduced Hilbert space. We will see that we can stay within solutions with $n = \gamma p$ and still obtain faces with negative area eigenvalues on the reduced Hilbert space.

\subsection{Master constraint}
\label{sec:master}
Compared with the solutions to the covariant simplicity constraints $F_1$ and $G_1$, the more interesting part follows now when we study the master constraint Eq.(\ref{eq:tsc20}) and how to solve it in the quantum theory,
\begin{equation}
\textbf{M} = \left(C_{\sl2c} - 2 \, Q_{\as11}\right) + \abs{\pi \omega}^2\,.
\label{eq:qu19}
\end{equation}

\noindent
Since we have already expressed this constraint in terms of the Casimirs, we only have to find a proper quantization of the last term. One finds \cite{twisted4} that the quantization of $\: \abs{\pi \omega}^2  \:$ should be given by
\begin{equation}
\hat{\omega}^A \hat{\pi}_A \hat{\bar{\pi}}_{\dot{B}} \hat{\bar{\omega}}^{\dot{B}} = - \omega^A \frac{\partial}{\partial \omega^A} \frac{\partial}{\partial \bar{\omega}^{\dot{B}}} \, \bar{\omega}^{\dot{B}}\,.
\label{eq:qu21}
\end{equation}

\noindent
Acting with Eq.(\ref{eq:qu21}) on a state, we get
\begin{align}
&- \omega^A \frac{\partial}{\partial \omega^A} \frac{\partial}{\partial \bar{\omega}^{\dot{B}}} \, \bar{\omega}^{\dot{B}} \, f^{(a,b)}\label{eq:qu22}\\[0.3\baselineskip]
&\quad =  - \omega^A \frac{\partial}{\partial \omega^A} \left(\bar{\omega}^{\dot{B}} \frac{\partial}{\partial \bar{\omega}^{\dot{B}}} + 2\right) \, f^{(a,b)} = - a (b+2) \, f^{(a,b)}\,,\notag
\end{align}

\noindent
where $-a (b+2)$ gives $(n^2 + 2n + 1 +p^2)$ when we use states in $(n,p)$ with non-negative $n$. If we use states from $(-n,-p)$, with $n \in \frac{\mathbb{N}_0}{2}$, then this gives $-a (b+2) = (n^2 - 2n + 1 +p^2)$. This distinction is important given our knowledge about the solutions of the area matching constraint Eq.(\ref{eq:quant16}). Now, what is the action of those two Casimirs on a general state $f^{(n,p)}$? The $\asl2c$ Casmir $C_{\sl2c} = C_2$ was given in Eq.(\ref{eq:quant6}) and gives
\begin{equation}
(\vec{L}^2 - \vec{K}^2) \, f^{(n,p)} = (n^2 - 1 -p^2) \, f^{(n,p)}\,,
\label{eq:qu23}
\end{equation}

\noindent
which, as we have already pointed out, is not sensitive to the change between $(n,p)$ and $(-n,-p)$, and the $\as11$ Casimir $Q_{\as11}$ gives with the plus convention
\begin{equation}
((L^3)^2 - (K^1)^2 - (K^2)^2) \, f^{(n,p)}_{j,m} =  j (j+1) \, f^{(n,p)}_{j,m}\,.
\label{eq:qu24}
\end{equation}

\noindent
One can show that this operator is also invariant with respect to the change between $(n,p)$ and $(-n,-p)$. Hence, we finally obtain
\begin{equation}
\widehat{\textbf{M}} \, f^{(n,p)}_{j,m} = \left[2n(n+1) - 2j(j+1)\right] \, f^{(n,p)}_{j,m} \stackrel{!}{=} 0
\label{eq:qu25}
\end{equation}

\noindent
and
\begin{equation}
\widehat{\textbf{M}} \, f^{(-n,-p)}_{j,m} = \left[2n(n-1) - 2j(j+1)\right] \, f^{(-n,-p)}_{j,m} \stackrel{!}{=} 0\,.
\label{eq:qu25.5}
\end{equation}

\noindent
In the standard time gauge, where the states $f^{(n,p)}_{j,m}$ diagonalize the $\asu2$ Casimir $\vec{L}^2$, the master constraint is solved by $n=j$. The solution with $n=-(j+1)$ does not occur in the decomposition Eq.(\ref{eq:quant18}). Even if we use that the representations $(n,p)$ and $(-n,-p)$ are unitarily equivalent, one finds that with $n = -n = j+1$ we have $j = n-1 < n$, which again does not occur in the decomposition Eq.(\ref{eq:quant18}), and hence $n=j$ is the only available solution. Now, in contrast to the $\su2$ case, the spectrum of $Q_{\as11}$ is determined by the four series $\mathcal{D}^{\pm}_k$ and $\mathcal{C}^{\varepsilon}_s$. Can the master constraint Eqs. (\ref{eq:qu25}) and (\ref{eq:qu25.5}) be solved with any of these states? Recall that for the principal series of the unitary irreducible representations of $\sl2c$ the parameter $n$ is an integer or half-integer. A \textit{priori} we can assume positive and negative values alike. But for $n(n \pm 1)$, there is a minimum value given by $-1/4$ for $n=-1/2$ or $n=1/2$. Otherwise, we have $n(n \pm 1) \geq 0 \:$ for all other $n$. Now, if we consider first the states of the two continuous series $\mathcal{C}^{\varepsilon}_s$ (with $\varepsilon \in \{0,\frac{1}{2}\}$), we see that Eqs. (\ref{eq:qu25}) and (\ref{eq:qu25.5}) with the plus convention for the $\as11$ Casimir $Q_{\as11}$ lead to
\begin{equation}
\left[n(n \pm 1) + \frac{1}{4} + s^2\right] \stackrel{!}{=} 0
\label{eq:quant100}
\end{equation}

\noindent
for both $\varepsilon$. It is clear that for most $n$ there is no solution to this condition. The only possible singular solution occurs for $n = \pm \frac{1}{2}$ and $\varepsilon = 0$, which is, however, of no relevance to us, since we consider $n \geq 0$ [even though we can solve Eq.(\ref{eq:qu25.5}) with $n = \frac{1}{2}$, this state will later be ruled out when solving the reduced area matching constraint]. Hence, for real $s \in \mathbb{R}_{\geq 0}$ we see that the master constraint cannot be solved by the states of the continuous series and the plus convention for $Q_{\as11}$. Note that this analysis transfers exactly to the other half-link in the $(\lambda,\sigma)$ variables.

Now, for the states of the discrete series $\mathcal{D}^{\pm}_k$, we obtain for Eq.(\ref{eq:qu25}) with $j=-k$
\begin{equation}
\left[n(n+1) - k(k-1)\right] \stackrel{!}{=} 0
\label{eq:quant101}
\end{equation}

\noindent
and see that the master constraint can be satisfied by the solutions
\begin{equation}
k = n+1 \qquad , \qquad k = -n\,.
\label{eq:quant102}
\end{equation}

\noindent
However, since we have $k \in \frac{\mathbb{N}}{2}$ and $n \in \frac{\mathbb{N}_0}{2}$, the second solution is not admissible. The first solution restricts furthermore the occurrence of the non-normalizable state $k = \frac{1}{2}$. For state with $(-n,-p)$, Eq.(\ref{eq:qu25.5}) gives with $j=-k$
\begin{equation}
\left[n(n-1) - k(k-1)\right] \stackrel{!}{=} 0\,,
\label{eq:quant101.8}
\end{equation}

\noindent
and we see that this is satisfied by the solutions
\begin{equation}
k = n \qquad , \qquad k = -n+1\,.
\label{eq:quant102.8}
\end{equation}

\noindent
Again, the second solution is not compatible with our range of parameter values. Using then the first solution in Eq.(\ref{eq:quant102}), we see that all the discrete states in $\mathcal{D}^{\pm}_k$ with $k \in \{1, \frac{3}{2}, 2, \cdots\}$ and $n \in \frac{\mathbb{N}_0}{2}$ solve the master constraint Eq.(\ref{eq:qu25}). For the first solution of Eq.(\ref{eq:quant102.8}), we see that $k,n \in \{\frac{1}{2}, 1, \frac{3}{2}, 2, \cdots\}$ solves the master constraint Eq.(\ref{eq:qu25.5}). However, we will see in the next section why it is preferable to change from the plus convention for $Q_{\as11}$ to the minus convention and to solve the master constraint using the continuous states instead.

\subsection{Reduced area matching constraint}
\label{sec:areareduced}
Now, we will consider the full reduced Hilbert space by imposing the reduced area matching constraint on the states that solve the simplicity constraints on the two half-links. From Eqs. (\ref{eq:quant16}) and (\ref{eq:quant17}), we learned that the area matching constraint imposes the conditions $n_t = -n_s$ and $p_t = -p_s$ on the tensor product states
\begin{equation}
f^{(n_s,p_s)}_{\text{left}} \otimes f^{(n_t,p_t)}_{\text{right}}\,.
\label{eq:quant103}
\end{equation}

\noindent
However, since we prefer to work with non-negative values for the $n_i$ labels we choose from the beginning states of the form
\begin{equation}
f^{(n_s,p_s)}_{\text{left}} \otimes f^{(-n_t,-p_t)}_{\text{right}}\,,
\label{eq:quant103.5}
\end{equation}

\noindent
which leads to the area matching condition $n_t = n_s \in \frac{\mathbb{N}_0}{2}$ and $p_t = p_s$. Since we already know from the simplicity constraints that $p_s = \gamma n_s$ or $p_s = - \frac{n_s}{\gamma}$ and similarly for the target half-link [which are not sensitive to a change between $(n,p)$ and $(-n,-p)$], we see that the area matching constraint reduces to only one condition, namely, $n_t = n_s$.

After imposing the master constraint on both half-links, we are left with the following possibilities on which we can impose the reduced area matching. First, we consider the case with $-j_s = k_s = n_s+1$ and $-j_t = k_t = n_t$. Solving the reduced area matching
\begin{equation}
\hat{C}_{\text{red}} \, \triangleright \left(f^{(n_s,p_s(n_s)),\pm}_{n_s+1,m_s} \otimes f^{(-n_t,-p_t(n_t)),\pm}_{n_t,m_t}\right) \stackrel{!}{=} 0
\label{eq:quant104}
\end{equation}

\noindent
leads to $n_t = n_s$ and hence both $n_i$ must be $n_i \in \frac{\mathbb{N}}{2}$. It furthermore implies $k_s = k_t+1$ and hence $k_s \in \{\frac{3}{2},2,\frac{5}{2},\cdots\}$ and $k_t \in \{\frac{1}{2},1,\frac{3}{2},\cdots\}$. From this, we obtain $K=k_s+k_t = 2n_s + 1$. Using now the decomposition Eq.(\ref{eq:repr4}), we find that we can obtain all the (integer) discrete states $\mathcal{D}^{\pm}_K$ with $K \geq 2$ as solutions to Eq.(\ref{eq:quant104}) from states satisfying the simplicity constraints. Explicitly, we have 
\begin{equation}
f^{(n_s,p_s(n_s)),\pm}_{n_s+1,m_s} \otimes f^{(-n_s,-p_t(-n_s)),\pm}_{n_s,m_t} = \bigoplus^{\infty}_{K = 2n_s+1} \mathcal{D}^{\pm}_K\,.
\label{eq:quant104.8}
\end{equation}

\noindent
Changing the order of the two states in the tensor product gives the same result. Now, let us consider the action of the reduced area matching operator on discrete states with opposite signs. Hence,
\begin{equation}
\hat{C}_{\text{red}} \, \triangleright \left(f^{(n_s,p_s(n_s)),\pm}_{n_s+1,m_s} \otimes f^{(-n_t,-p_t(n_t)),\mp}_{n_t,m_t}\right) \stackrel{!}{=} 0\,.
\label{eq:quant105}
\end{equation}

\noindent
Using again the solution $n_t = n_s$, we find that $k_s + k_t = 2n_s + 1 \in \mathbb{Z}$, and hence for the decomposition Eq.(\ref{eq:repr5}), we get $K_{\text{min}} = 1$ and $\varepsilon = 0$. Furthermore, we have $k_s - k_t = 1$ and $k_t - k_s = -1$, and hence one finds that those states that satisfy the simplicity constraints and the reduced area matching are given by
\begin{equation}
f^{(n_s,p_s(n_s)),\pm}_{n_s+1,m_s} \otimes f^{(-n_s,-p_t(n_s)),\mp}_{n_s,m_t} = \mathcal{D}^{\pm}_1 \: \oplus \int^{\infty \oplus}_0{\mathcal{C}^{0}_s \: ds}\,.
\label{eq:quant107}
\end{equation}

\noindent
Hence, we see that we do not obtain all the states we need to span $\s11$ spin networks, i.e., all the states that appear in the harmonic analysis of functions on $\s11$. We only obtain the discrete states $\mathcal{D}^{\pm}_K$ with $K \in \mathbb{N}$ and are missing all the half-integral values $K \in \frac{\mathbb{N}}{2}$. Similarly, we only obtain the even continuous states $\mathcal{C}^{0}_s$, but we are missing the odd states with $\varepsilon = \frac{1}{2}$. This is a result of the reduced area matching constraint, which does not allow for tensor-product states that have integer labels on the left factor and half-integer labels on the right factor (or vice versa). Hence, in the decomposition, only states with integer labels and/or states with $\varepsilon = 0$ appear. However, this problem can be solved as follows. The requirement that we need all unitary irreducible Plancherel representations of $\s11$ forces us to choose the minus convention in Eq.(\ref{eq:repr1}). This gives for the master constraint now the conditions
\begin{equation}
\widehat{\textbf{M}} \, f^{(\pm n,\pm p)}_{j,m} = \left[2n(n \pm 1) + 2j(j+1)\right] \, f^{(\pm n, \pm p)}_{j,m} \stackrel{!}{=} 0\,,
\label{eq:quant120}
\end{equation}

\noindent
which can now not be satisfied by the states of the discrete series anymore but by the states of the continuous series. For the states $f^{(\pm n,\pm p)}_{s,m}$, one obtains the solution
\begin{equation}
s^{\pm}(n)=\frac{\sqrt{(2n \pm 1)^2 - 2}}{2}\,.
\label{eq:quant121}
\end{equation}

\noindent
For the states $f^{(n,p)}_{s,m}$, this is strictly positive for $n \in \frac{\mathbb{N}}{2}$, hence $n=0$ is ruled out, and for the states $f^{(-n,-p)}_{s,m}$, we have to restrict $n$ such that $n \in \{\frac{3}{2},2,\frac{5}{2},\cdots\}$. The reason why we can now use those states to obtain the full reduced Hilbert space is that neither the simplicity constraints nor the reduced area matching constraint restricts the labels $\varepsilon_s$ and $\varepsilon_t$, which, according to Eq.(\ref{eq:repr6}), determine which states appear in the decomposition, i.e., $K_{\text{min}}$ and $\varepsilon$ are now determined by $\varepsilon_s + \varepsilon_t$, which can now be freely chosen to be integral or half-integral. Explicitly, we find that the simplicity and reduced area matching constraints are now solved by the states
\begin{equation}
\Psi^{n_s, \varepsilon_s, \varepsilon_t}_{m_s,m_t} \equiv f^{(n_s,p_s(n_s)),\varepsilon_s}_{s^+_1(n_s),m_s} \otimes f^{(-n_s,-p_t(n_s)),\varepsilon_t}_{s^-_2(n_s),m_t}\,,
\label{eq:quant122}
\end{equation}

\noindent
where now $n_s \geq \frac{3}{2}$. Again, we can now freely choose whether $\varepsilon_s + \varepsilon_t$ is integral, which gives from Eq.(\ref{eq:repr6}) the states
\begin{equation}
\bigoplus^{\infty}_{K=1} \mathcal{D}^{+}_K \: \oplus \bigoplus^{\infty}_{K=1} \mathcal{D}^{-}_K \: \oplus 2 \int^{\infty \oplus}_0{\mathcal{C}^0_s \: ds}\,,
\label{eq:quant123}
\end{equation}

\noindent
or whether $\varepsilon_s + \varepsilon_t$ is half-integral, which gives the states
\begin{equation}
\bigoplus^{\infty}_{K=\frac{3}{2}} \mathcal{D}^{+}_K \: \oplus \bigoplus^{\infty}_{K=\frac{3}{2}} \mathcal{D}^{-}_K \: \oplus 2 \int^{\infty \oplus}_0{\mathcal{C}^{\frac{1}{2}}_s \: ds}
\label{eq:quant124}
\end{equation}

\noindent
and thus we see that we obtain all the discrete states with $K \in \{1, \frac{3}{2}, 2, \frac{5}{2}, \cdots \}$ as well as all the continuous states spanning our reduced Hilbert space. Note that, due to the integral over the continuous parameter $s$ in both decompositions Eqs. (\ref{eq:quant123}) and (\ref{eq:quant124}), we obtain all continuous states for arbitrary $s \in \mathbb{R}_{\geq 0}$ in the coupled basis and not just those that satisfy the discreteness constraint Eq.(\ref{eq:quant121}). This can be seen explicitly by considering the Clebsch-Gordan coefficients of the above decompositions. Even when both states in the decoupled basis satisfy the condition Eq.(\ref{eq:quant121}), one obtains nonzero Clebsch-Gordan coefficients for general $s \in \mathbb{R}_{\geq 0}$ in the coupled basis. This means in particular that the reduced Hilbert space includes indeed all the necessary $\s11$ Plancherel representations that are necessary to expand states in the holonomy representation, i.e., certain $\mathbb{C}$-valued functions on $\s11$, in terms of a spin network basis. Thus, this gives perfect agreement of our reduced Hilbert space and the quantization of 3D Lorentzian gravity \cite{freidel3d, sellaroli3}. Note, that for such spin network states we can obtain links that are labeled by arbitrary continuous states with $s \in \mathbb{R}_{\geq 0}$. On the level of the coupled basis of the reduced Hilbert space, one then finds that the area associated with such links can be continuous, again in agreement with the 3D Lorentzian case. However, those states are not physical, in the sense that they do not satisfy simplicity constraints and area matching, i.e., they are not of the form Eq.(\ref{eq:quant122}). If we consider a general $\s11$ spin network state, which is labeled by continuous $s$ values, we know from the inverse decompositions of Eqs. (\ref{eq:quant123}) and (\ref{eq:quant124}) how to embed those states into our solution space of simplicity and area matching constraints via Eqs. (\ref{eq:livmap}) and (\ref{eq:livmap2}). This is basically the Lorentzian version of the Livine-Dupuis map known from the standard EPRL model and shows nicely how to embed the three-dimensional Lorentzian Ponzano-Regge model into our four-dimensional setting. This gives, furthermore, an explicitly mechanism that shows how we can have continuous eigenvalues on the 3D level, but when we embed those states into the solution space of simplicity and area matching constraint those eigenvalues become strictly discrete. Note that one does not need this decomposition explicitly to calculate, for example, the area operator eigenvalues of the state Eq.(\ref{eq:quant122}) as we will see in the next section. We consider it another important result of our work that we obtain a reduced Hilbert space with enough states such that one obtains a valid $\s11$ spin network decomposition. Compared with the standard time gauge case, where one solves both simplicity constraints on each half-link and obtains already all the necessary $\su2$ states on each half-link (which are then glued using the area matching), it was necessary in our case to understand that, even though we just obtain a subclass of representations per half-link as solutions to the simplicity constraints, all the required $\s11$ states arise after the decomposition of the tensor product states and imposition of the reduced area matching.

\subsection{Area spectra}
\label{sec:area}
In Lorentzian spin foam models \cite{timelikefaces1,timelikefaces2,timelikefaces3,LorentzianSFM1,LorentzianSFM2} and LQG, there are two major issues concerning the spectra of geometrical operators and the area operator in particular. The first is about the question of whether those operators have discrete or continuous spectra \cite{lengthandareaspec,lengthandareaspec2,lengthandareaspec3}, and the second concerns the appearance of the Barbero-Immirzi parameter \cite{geiller2}. The first problem can, at least in four dimensions, be further separated into whether we are talking about spectra on the kinematical level or at the level of the physical Hilbert space; see, for example, Refs. \cite{spectradiscrete, rovrep}. 

In LQG, the area operator is essentially given by (the square root of) the $\asu2$ Casimir since the (densitized) flux operators satisfy a $\asu2$ algebra and thus the quantization of the classical expression for the area (squared) leads explicitly to $\vec{L}^2$ (with a $\gamma$-dependent prefactor), \cite{rovelli,thiemann}. This leads then to the discrete spectra for the area (on the kinematical Hilbert space). However, there have been other proposals for the area operator within covariant formulations of LQG \cite{lengthandareaspec2,timelikefaces3} that potentially lead to continuous and $\gamma$-independent area spectra. That there are cases when the Barbero-Immirzi parameter disappears from the area spectra was also observed in Ref. \cite{geiller2} and is a result we will discuss in this section using our twistorial description. Our definition of the area operator was given in Eq.(\ref{eq:areadef1}) by the Plebanski 2-form $\Sigma$, and we consider
\begin{equation}
\hat{\mathcal{A}}^2 \equiv \frac{1}{2} \hat{\Sigma}_{IJ} \hat{\Sigma}^{IJ}\,.
\label{eq:area1}
\end{equation}

\noindent
Using the vector representation in terms of rotation and boost generators allows us to understand its reduction classically as follows. Recall that we have associated the $\asl2c$ generators with $B^{IJ}$ as in Eq.(\ref{eq:bfield1}). Furthermore, we have
\begin{equation}
\Sigma = - \frac{\gamma^2}{1+\gamma^2} \left(\ast + \frac{1}{\gamma}\right) B\,,
\label{eq:area2}
\end{equation}

\noindent
which, together with Eq.(\ref{eq:area1}), gives
\begin{align}
\mathcal{A}^2 &= \frac{\gamma^4}{2 (1+\gamma^2)^2} \left((\ast B) + \frac{B}{\gamma}\right)_{IJ} \left((\ast B) + \frac{B}{\gamma}\right)^{IJ}\label{eq:area3}\\[0.3\baselineskip]
&= \frac{\gamma^4}{(1+\gamma^2)^2} \left(\left(\frac{1}{\gamma^2} - 1\right)(\vec{L}^2 - \vec{K}^2) + \frac{1}{\gamma} (\ast B)_{IJ} B^{IJ}\right)\,.\notag
\end{align}

\noindent
Using that
\begin{equation}
(\ast B)_{IJ} B^{IJ} = -4 \left(L^1 K^1 + L^2 K^2 + L^3 K^3\right)
\label{eq:area4}
\end{equation}

\noindent
we get with the simplicity constraints $\Sigma^{3i} = 0$, i.e.,
\begin{equation}
K^3 = - \gamma L^3 \quad , \quad L^1 = \gamma K^1 \quad , \quad L^2 = \gamma K^2\,,
\label{eq:area5}
\end{equation}

\noindent
that
\begin{align}
\vec{L}^2 - \vec{K}^2 &= (1 - \gamma^2) \, Q_{\as11}\,,\label{eq:area6}\\[0.3\baselineskip]
(\ast B)_{IJ} B^{IJ} &= 4 \gamma \, Q_{\as11}\,,
\label{eq:area6.5}
\end{align}

\noindent
which finally leads to
\begin{equation}
\mathcal{A}^2 = \gamma^2 \, Q_{\as11}\,.
\label{eq:area7}
\end{equation}

\noindent
Now, if we use the dual simplicity constraints $(\ast \Sigma)^{3i} = 0$, or
\begin{equation}
K^3 = \frac{1}{\gamma} L^3 \quad , \quad L^1 = -\frac{1}{\gamma} K^1 \quad , \quad L^2 = -\frac{1}{\gamma} K^2\,,
\label{eq:area8}
\end{equation}

\noindent
we obtain instead
\begin{align}
\vec{L}^2 - \vec{K}^2 &= \left(1 - \frac{1}{\gamma^2}\right) \, Q_{\as11}\,,\label{eq:area9}\\[0.3\baselineskip]
(\ast B)_{IJ} B^{IJ} &= - \frac{4}{\gamma} \, Q_{\as11}
\label{eq:area9.5}
\end{align}

\noindent
and hence
\begin{equation}
\mathcal{A}^2 = - Q_{\as11}\,.
\label{eq:area10}
\end{equation}

\noindent
This already indicates that the Barbero-Immirzi parameter $\gamma$ seems to disappear in the spectrum for states that solve the dual simplicity constraints $(\ast \Sigma)^{3i} = 0$, similarly to the results found in Ref. \cite{geiller2}\footnote{However, note that the same reasoning works for the $\su2$ case, where we can equally consider $\Sigma^{0i} = 0$ or the dual $(\ast \Sigma)^{0i} = 0$ but with a timelike normal vector $N^I$, and we still obtain that in the first case we have a $\gamma$ dependence, i.e., $\mathcal{A}^2_{\su2} = \gamma^2 \, \vec{L}^2$, and in the other case, we have a sign flip, and $\gamma$ disappears, i.e., $\mathcal{A}^2_{\su2} = - \vec{L}^2$.}. Now, let us consider the quantized area operator in the twistorial parametrization. Using the action of $\widehat{\pi \omega}$ and $\widehat{\bar{\pi} \bar{\omega}}$ on the homogeneous functions $f^{(a,b)} \in \mathcal{H}^{(a,b)}$
\begin{equation}
\widehat{\pi \omega} \triangleright f^{(a,b)} = -i \hbar \, [a+1] \: f^{(a,b)}
\label{eq:area11}
\end{equation}

\noindent
and 
\begin{equation}
\widehat{\bar{\pi} \bar{\omega}} \triangleright f^{(a,b)} = -i \hbar \, [b+1] \: f^{(a,b)}\,,
\label{eq:area12}
\end{equation}

\noindent
we obtain with Eqs. (\ref{eq:areadef1}) and (\ref{eq:quant5}) that
\begin{align}
&\hat{\mathcal{A}}^2 \triangleright f^{(a,b)} = \frac{\gamma^2}{8} \left(\frac{\widehat{\pi \omega} \, \widehat{\pi \omega}}{(\gamma + i)^2} + \frac{\widehat{\bar{\pi} \bar{\omega}} \, \widehat{\bar{\pi} \bar{\omega}}}{(\gamma - i)^2}\right) \triangleright f^{(a,b)}\notag\\[0.3\baselineskip]
&= -\frac{\hbar^2}{8} \frac{\gamma^2}{(\gamma^2 + 1)^2} \left[(\gamma^2 - 1) (a^2 + b^2 + 2a + 2 b + 2)\right.\notag\\[0.3\baselineskip]
&\qquad \qquad \qquad \qquad \left.- 2 i \gamma (a^2 - b^2 + 2a - 2b)\right] f^{(a,b)}\notag\\[0.3\baselineskip]
&= -\frac{\hbar^2}{4} \frac{\gamma^2}{(\gamma^2 + 1)^2} \left[(\gamma^2 - 1) (n^2 - p^2) - 4 \gamma n p\right] f^{(a,b)}\,.
\label{eq:area13}
\end{align}

\noindent
Now, if we consider the solutions to the simplicity constraints, $p = \gamma n$ for $F_1 = 0$ and $p = - n / \gamma$ for $G_1 = 0$, we obtain
\begin{equation}
\hat{\mathcal{A}}^2 \triangleright f^{(n,\gamma n)} = \frac{\hbar^2}{4} \gamma ^2 n^2 \: f^{(n,\gamma n)}
\label{eq:area14}
\end{equation}

\noindent
and
\begin{equation}
\hat{\mathcal{A}}^2 \triangleright f^{(n,- n / \gamma)} = -\frac{\hbar^2}{4} n^2 \: f^{(n,- n / \gamma)}
\label{eq:area15}
\end{equation}

\noindent
respectively. First, note that we find that, indeed, the area eigenvalues switch sign between the two branches with $p = \gamma n$ and $p = - n / \gamma$, respectively. Hence, our identification of the constraints $(F_1,F_2)$ with the spacelike case and the constraints $(G_1,G_2)$ with the timelike case seems justified. Furthermore, we again confirm that the area spectrum for timelike faces seems to not depend on $\gamma$. Second, note the different nature between Eqs. (\ref{eq:area14}) and (\ref{eq:area15}) on the one hand and Eqs. (\ref{eq:area7}) and (\ref{eq:area10}) on the other. For the calculation in Eqs. (\ref{eq:area14}) and (\ref{eq:area15}), we have used the covariant version of the area operator Eq.(\ref{eq:area1}) and then imposed the solutions of the simplicity constraints on the area eigenvalues, which leads us, in the spacelike case as well as in the timelike case, to discrete area eigenvalues, which is in contrast to the statement often made in the literature, e.g, Refs. \cite{geiller2, freidel3d, covlqg2, spectradiscrete}, that in Lorentzian models we have necessarily continuous spectra, due to the noncompactness of the gauge group. In the formulas leading to Eqs. (\ref{eq:area7}) and (\ref{eq:area10}), on the other hand, we have first reduced the operator by the simplicity constraints. If we use now for the (reduced) area operators Eqs. (\ref{eq:area7}) and (\ref{eq:area10}) instead, we first would notice that this operator does not act on the covariant labels $(a,b)$ but on the $\s11$ labels $j(k)$ and $j(s)$. In this situation, one might wonder whether we actually recover continuous spectra for the continuous states with $j(s)$ and $Q^c_{\s11} = -j(j+1) = \frac{1}{4} + s^2$, which is related to our discussion about whether we have all the continuous states available in the reduced Hilbert space or just a discrete subset. We will see now that both ways, reducing the eigenvalues of the covariant area operator or first reducing the area operator, are consistent and lead in both cases to a discrete area eigenvalue spectrum for those states that solve the area matching and simplicity constraints. Consider first a state of the form Eq.(\ref{eq:quant122}) with $p_s = \gamma n_s = p_t$. Then, $Q_{\as11}$ acts as
\begin{align}
&Q_{\as11} \triangleright \left(f^{(n_s,\gamma n_s),\varepsilon_s}_{s^+_1(n_s),m_s} \otimes f^{(-n_s,-\gamma n_s),\varepsilon_t}_{s^-_2(n_s),m_t}\right)\notag\\[0.3\baselineskip]
&= \left(Q_{\as11} \triangleright f^{(n_s,\gamma n_s),\varepsilon_s}_{s^+_1(n_s),m_s}\right) \otimes f^{(-n_s,-\gamma n_s),\varepsilon_t}_{s^-_2(n_s),m_t}\notag\\[0.3\baselineskip]
& \qquad \qquad + \: f^{(n_s,\gamma n_s),\varepsilon_s}_{s^+_1(n_s),m_s} \otimes \left(Q_{\as11} \triangleright f^{(-n_s,-\gamma n_s),\varepsilon_t}_{s^-_2(n_s),m_t}\right)\notag\\[0.3\baselineskip]
&= \left(\frac{1}{4} + (s^+_1(n_s))^2 + \frac{1}{4} + (s^-_2(n_s))^2\right)\notag\\[0.3\baselineskip]
& \qquad \qquad \times f^{(n_s,\gamma n_s),\varepsilon_s}_{s^+_1(n_s),m_s} \otimes f^{(-n_s,-\gamma n_s),\varepsilon_t}_{s^-_2(n_s),m_t}\notag\\[0.3\baselineskip]
&= 2 n^2_s \: \: f^{(n_s,\gamma n_s),\varepsilon_s}_{s^+_1(n_s),m_s} \otimes f^{(-n_s,-\gamma n_s),\varepsilon_t}_{s^-_2(n_s),m_t}
\label{eq:area150}
\end{align}

\noindent
and hence with $\hat{\mathcal{A}}^2 = \gamma^2 \, Q_{\as11}$, we get
\begin{align}
&\hat{\mathcal{A}}^2 \: \triangleright \: \left(f^{(n_s,\gamma n_s),\varepsilon_s}_{s^+_1(n_s),m_s} \otimes f^{(-n_s,-\gamma n_s),\varepsilon_t}_{s^-_2(n_s),m_t}\right) =\label{eq:area151}\\[0.3\baselineskip]
&\qquad \qquad \quad 2 \gamma^2 n^2_s \: \: \left(f^{(n_s,\gamma n_s),\varepsilon_s}_{s^+_1(n_s),m_s} \otimes f^{(-n_s,-\gamma n_s),\varepsilon_t}_{s^-_2(n_s),m_t}\right)\,.\notag
\end{align}

\noindent
Comparing this with Eq.(\ref{eq:area14}), where the missing factor of $\hbar^2$ is included in $Q_{\as11}$ and up to an irrelevant factor of $\frac{1}{8}$, we showed the consistency between the two ways of obtaining the area eigenvalues. If we consider now similarly the dual case with $p_s = -\frac{n_s}{\gamma} = p_t$, we have $\hat{\mathcal{A}}^2 = - Q_{\as11}$, cf. Eq.(\ref{eq:area10}), and we obtain instead
\begin{align}
&\hat{\mathcal{A}}^2 \: \triangleright \: \left(f^{(n_s,-\frac{n_s}{\gamma}),\varepsilon_s}_{s^+_1(n_s),m_s} \otimes f^{(-n_s,\frac{n_s}{\gamma}),\varepsilon_t}_{s^-_2(n_s),m_t}\right) =\label{eq:area152}\\[0.3\baselineskip]
&\qquad \qquad \quad -2 n^2_s \: \: \left(f^{(n_s,-\frac{n_s}{\gamma}),\varepsilon_s}_{s^+_1(n_s),m_s} \otimes f^{(-n_s,\frac{n_s}{\gamma}),\varepsilon_t}_{s^-_2(n_s),m_t}\right)\,.\notag
\end{align}

\noindent
This matches the result of Eq.(\ref{eq:area15}) and $\gamma$ seems to not appear. Note that, due to the area matching constraint, we must have $p_t = p_s$. Hence, if we were to consider the coupling of states with $p_s = \gamma n_s$ and $p_t = - \frac{n_s}{\gamma}$, or vice versa, the condition $p_t = p_s$ leads to the requirement that $\gamma$ must be imaginary, i.e., $\gamma = \pm i$, which might be related to the self-dual Ashtekar variables that have recently been investigated in Refs. \cite{sd1, sd2, sd3}. It is tempting to interpret this in some way as a coupling of a spacelike state on one side of the link with a timelike state on the other side. However, throughout this work, we have assumed real $\gamma$, and hence considering complex $\gamma$ is merely a speculation at this level. Furthermore, it is important to note that in the theory as presented in this paper taking $\gamma$ to be complex would take us out of the unitary representations of $\sl2c$.

If we want to avoid using the dual constraints $p = - \frac{n}{\gamma}$, because spacelike states as well as timelike states should be included already in just the case with $p = \gamma n$, we can consider the explicit decomposition of the solution state Eq.(\ref{eq:quant122}) into its irreducible components following Eqs. (\ref{eq:quant123}) and (\ref{eq:quant124}). Acting with $Q_{\as11}$ onto those irreducible states will give positive as well as negative eigenvalues of the continuous series and the discrete series, respectively. Hence, in this picture, the timelike states are associated with the discrete series states, which are composed as the tensor product of two continuous states. In the reversed direction, imagine we have a spin network decorated with $\s11$ representations $j(k)$ or $j(s)$; then, we can think of a generalized Livine-Dupuis map\footnote{In the $\su2$ case, the Livine-Dupuis map embeds the $\su2$ representations into the subspace of the canonical basis that satisfies the simplicity constraints as $\ket{j,m} \hookrightarrow \ket{(j,\gamma j),j,m}$.}, which maps the states of the $\s11$ spin network into the solution states of the area matching and simplicity constraint as
\begin{equation}
\ket{j(k),m} \mapsto \sum_{m_s, m_t} C(n_s) f^{(n_s,p_s(n_s)),\varepsilon_s}_{s^+_1(n_s),m_s} \otimes f^{(-n_s,-p_t(n_s)),\varepsilon_t}_{s^-_2(n_s),m_t}\,,
\label{eq:livmap}
\end{equation}

\noindent
or for the continuous states with $j(s)$ as
\begin{equation}
\ket{j(s),m} \mapsto \sum_{m_s, m_t} \tilde{C}(n_s) f^{(n_s,p_s(n_s)),\varepsilon_s}_{s^+_1(n_s),m_s} \otimes f^{(-n_s,-p_t(n_s)),\varepsilon_t}_{s^-_2(n_s),m_t}\,,
\label{eq:livmap2}
\end{equation}

\noindent
where $C(n_s)$ and $\tilde{C}(n_s)$ depend besides $n_s$ on $k$ or $s$ and on $m_s, m_t$ and denote the Clebsch-Gordan coefficients corresponding to the inverse of the decompositions in Eqs. (\ref{eq:quant123}) and (\ref{eq:quant124}). The details of this embedding will be relevant for the construction of a generalized spin foam model, and thus we will leave them for future investigations.

Finally, let us comment again on the fate of the Barbero-Immirzi parameter. We point out that we discuss here only the appearance of $\gamma$ in the eigenvalues of the area operator for timelike faces and not whether the physical Hilbert space will be $\gamma$ dependent or not. From Eqs. (\ref{eq:area10}) and (\ref{eq:area15}), with the solution of the dual simplicity constraints $(\ast \Sigma)^{3i} = 0$, i.e., $p = - n / \gamma$, we confirmed the statement that was made in Refs. \cite{geiller1} and \cite{geiller2} that the spectrum of timelike faces does not depend on $\gamma$. However, there is a possibility that $\gamma$ might actually reappear as follows. Note that when we introduce dimensionful constants the area operator $\sqrt{\hat{\mathcal{A}}^2}$ has eigenvalues proportional to the Planck length \cite{rovelli}; i.e., for the standard $\su2$ case, we have
\begin{equation}
\hat{\mathcal{A}} \triangleright \ket{j} = 8 \pi \gamma l^2_{P} \sqrt{j(j+1)} \: \ket{j}\,,
\label{eq:area200}
\end{equation}

with $l^2_{P} = \hbar G / c^3$, and hence we see that it depends on the gravitational constant $G$. This certainly holds true for the spacelike faces and the space-gauge simplicity constraints $(F_1,F_2)$. If we consider now the area spectrum of timelike faces, we would assume that it is proportional to either $t_{P} l_{P}$ or $t^2_P$, where $t_P$ is the Planck time with $t^2_P = l^2_P / c^2$. In either case, we again find that the spectrum is proportional to $G$. However, if we go back to the original Holst action we started with in Eq.(\ref{eq:qu38}) and note that there is a prefactor of $1 / (16 \pi G)$, then we notice that the dual simplicity constraints  $(\ast \Sigma)^{3i} = 0$, i.e., $(G_1,G_2)$, lead to Einstein-Cartan gravity with the dual Barbero-Immirzi parameter $\tilde{\gamma} = - 1 / \gamma$ and a scaled gravitational constant $\tilde{G} = G \gamma$. Now, in this situation, it appears as if $\gamma$ does not appear in the area operator, but, in fact, if we consider the proportionality with $\tilde{G} = G \gamma$, we see that it still appears via the rescaling of $G$. Following this reasoning would imply that all our area spectra are linearly dependent on $\gamma$ as in the standard $\su2$ case.

\section{Discussion}
\label{sec:disc}
We introduced and investigated in this paper the notion of timelike twisted geometries. Together with the standard time-gauge case \cite{twisted3, twisted4}, which leads to $\su2$ spin networks, and the more recently introduced null twisted geometries \cite{nulltwisted}, this completes the application of the twistorial variables to all types of Lorentzian geometries. We showed in the classical setting explicitly how the simplicity constraints with a spacelike normal vector reduce $\bsl2c$ to $\bs11$ on each link and similarly, how in the quantum theory the reduced Hilbert space is spanned by $\s11$ spin networks. Our results fit nicely with the recent spinorial investigations of 3D Lorentzian gravity in Ref. \cite{sellaroli3} and can be seen as giving an independent derivation of (some of) those results from a four-dimensional perspective.

We furthermore confirmed results from Refs. \cite{geiller1} and \cite{geiller2} concerning the fate of the Barbero-Immirzi parameter but provide a different interpretation, namely, that $\gamma$ still enters the spectrum of the area operator when we take a rescaling of the gravitational constant $G$ into account. We further discussed the nature of the eigenvalues of the area operator and why they turn out to be discrete for spacelike faces and timelike faces alike, despite the underlying noncompact gauge group. This is a result of the simplicity constraints that provide relations between continuous and discrete representation labels, and hence no continuous spectra appear for the states in Eq.(\ref{eq:quant122}), which satisfy the simplicity constraints and area matching. This might be interpreted as saying that in LQG and spin foams not only lengths and areas but also time intervals are discrete with a minimal nonzero value. An open question concerns the problem of imposing the constraints in a different order than the one chosen by us. It seems to us not obvious at the moment, how to obtain the full reduced Hilbert space of $\s11$ Plancherel representations when one first imposes the full area matching constraint and then tries to impose the simplicity constraints, since in this order the master constraint always rules out either the discrete states or the continuous states.

The main result of this paper, however, is the derivation of the quantum states that correspond to quantum timelike 2-surfaces in terms of spinorial variables. The spinor variables have proven very useful in the past for the asymptotic analysis of the standard EPRL model. This opens the door for further investigations of such generalized spin foam models as proposed in Refs. \cite{LorentzianSFM1} and \cite{LorentzianSFM2}. The most pressing question is certainly whether such generalized models that include timelike components at least share or maybe even improve the semiclassical limit of the EPRL model. One should also investigate possible connections with the proper EPRL vertex amplitude of Refs. \cite{proper1} and \cite{proper2}. Furthermore, it seems now possible to use our variables to consider the model proposed in Ref. \cite{simplicial} in the Lorentzian setting. Further possible research directions concern a more detailed investigation of the alternative set of constraints mentioned in footnote 2 as well as the question of a more in depth study of Lorentzian intertwiner spaces that arise from the coupling of several $\s11$ representations. In that regard it is interesting to consider, for example, the Lorentzian generalization of the Livine-Speziale coherent states and how they relate explicitly to the classical Lorentzian phase space of shapes underlying the $\s11$ intertwiner spaces. This work will appear elsewhere.

\begin{center}
\textbf{Acknowledgments}
\end{center}
The author would like to thank S. Speziale, W.M. Wieland, F. Girelli, and G. Sellaroli for numerous discussions about this project and helpful comments while preparing this manuscript. This research was supported by an Ontario Trillium Scholarship.



\bibliography{references}
\bibliographystyle{apsrev4-1}

\end{document}